\documentclass[reprint,superscriptaddress,noeprint,amssymb,amsmath,aps,prx,nolongbibliography]{revtex4-2}

\setcitestyle{numbers,square,super}
\usepackage{minitoc}

\usepackage{graphicx}% Include figure files
\usepackage{dcolumn}% Align table columns on decimal point
\usepackage{bm}% bold math
\usepackage{siunitx}
\usepackage{booktabs}
\usepackage[usenames,dvipsnames]{xcolor}
\usepackage{tcolorbox}
\usepackage{tabularx}
\usepackage{array}
\usepackage{colortbl}
\usepackage{braket}
\usepackage{gensymb}
\usepackage{pdfpages}
\usepackage{amsmath}
\usepackage{mathrsfs}
\usepackage{blindtext}
\usepackage{minitoc}
\usepackage{dsfont}
\usepackage{multirow}
\usepackage{lipsum}  
\usepackage{float}

\usepackage{bibunits} 
\defaultbibliographystyle{naturemag} 
\defaultbibliography{main}      

\usepackage{amssymb}

\makeatletter
\def\maketitle{
\@author@finish
\title@column\titleblock@produce
\suppressfloats[t]}
\makeatother

\newcommand\SupplementaryMaterials{%
  \xdef\presupfigures{\arabic{figure}}% save the current figure number
  \xdef\presupsections{\arabic{section}}% save the current section number
  \renewcommand\thefigure{S\fpeval{\arabic{figure}-\presupfigures}}
  \renewcommand\thesection{S\fpeval{\arabic{section}-\presupsections}}
}

\makeatletter
\patchcmd{\@outputpage@head}{\@ifx{\LS@rot\@undefined}{}{\LS@rot}}{}{}{}
\makeatother

\tcbuselibrary{skins}

\tcbset{tab2/.style={colback=gray!5!white,colframe=black!50!black,colbacktitle=Gray!40!white,
coltitle=black,center title}}

\begin{document}

\title{Quantum critical electro-optic and piezo-electric nonlinearities}

\author{Christopher P. Anderson$^\dagger$}
\thanks{These authors contributed equally to this work}
\affiliation{E. L. Ginzton Laboratory, Stanford University, Stanford, California 94305, USA}
\affiliation{Department of Materials Science and Engineering, University of Illinois Urbana-Champaign, Urbana, IL 61801, USA}
\email{cpand@illinois.edu, jela@stanford.edu}

\author{Giovanni Scuri}
\thanks{These authors contributed equally to this work}
\affiliation{E. L. Ginzton Laboratory, Stanford University, Stanford, California 94305, USA}

\author{Aaron Chan}
\affiliation{Department of Physics, University of Michigan, Ann Arbor 48109, USA}

\author{Sungjun Eun}
\affiliation{E. L. Ginzton Laboratory, Stanford University, Stanford, California 94305, USA}

\author{Alexander D. White}
\affiliation{E. L. Ginzton Laboratory, Stanford University, Stanford, California 94305, USA}

\author{Geun Ho Ahn}
\affiliation{E. L. Ginzton Laboratory, Stanford University, Stanford, California 94305, USA}

\author{Christine Jilly}
\affiliation{Stanford Nano Shared Facilities, Stanford University, Stanford, California 94305, United States}

\author{Amir Safavi-Naeini}
\affiliation{E. L. Ginzton Laboratory, Stanford University, Stanford, California 94305, USA}

\author{Kasper Van Gasse}
\affiliation{E. L. Ginzton Laboratory, Stanford University, Stanford, California 94305, USA}
\affiliation{Ghent University, 9052 Gent, Belgium}

\author{Lu Li}
\affiliation{Department of Physics, University of Michigan, Ann Arbor 48109, USA}

\author{Jelena Vu\v{c}kovi\'{c}$^\dagger$}
\affiliation{E. L. Ginzton Laboratory, Stanford University, Stanford, California 94305, USA}
\email{jela@stanford.edu}

\date{\today}
\begin{bibunit} 
\begin{abstract}

Electro-optics, the tuning of optical properties of materials with electric fields, is key to a multitude of quantum and classical photonics applications\cite{Sinatkas2021}. However, a major obstacle preventing many emerging use cases is inefficient modulation in cryogenic environments, as traditional tuning mechanisms degrade at low temperatures\cite{Pintus2022,Eltes2020}. Guided by the connection between phase transitions and nonlinearity, we identify the quantum paraelectric perovskite SrTiO$_3$ (STO) as the strongest cryogenic electro-optic photonic material. As a result of the unique quantum paraelectric phase of STO, we demonstrate a dynamically tunable linear Pockels coefficient ($r_{33}$) exceeding 500 pm/V at $T=5$ K, and study its full temperature and bias dependence. We also measure an enhanced piezo-electric coefficient ($d_{33}$) above 90 pC/N. Both of these coefficients exceed all previously reported values for cryogenic materials, including lithium niobate ($r_{33}\approx24$~pm/V) and barium titanate ($r_{42}\approx170$~pm/V)\cite{10.1063/5.0192018, Eltes2020, LN_lowtemp}. Furthermore, by tuning STO towards \textit{quantum criticality} with oxygen isotope substitution\cite{itoh} we more than double the optical and piezo-electric nonlinearities, demonstrating a linear Pockels coefficient above 1100 pm/V. Our results probe the link between quantum phase transitions, dielectric susceptibility, and optical nonlinearities, unlocking opportunities in cryogenic optical and mechanical systems, and provide a framework for discovering new nonlinear materials.

\end{abstract}
\maketitle

\begin{figure}[htb!] 
\begin{center}
\includegraphics[width=1\columnwidth]{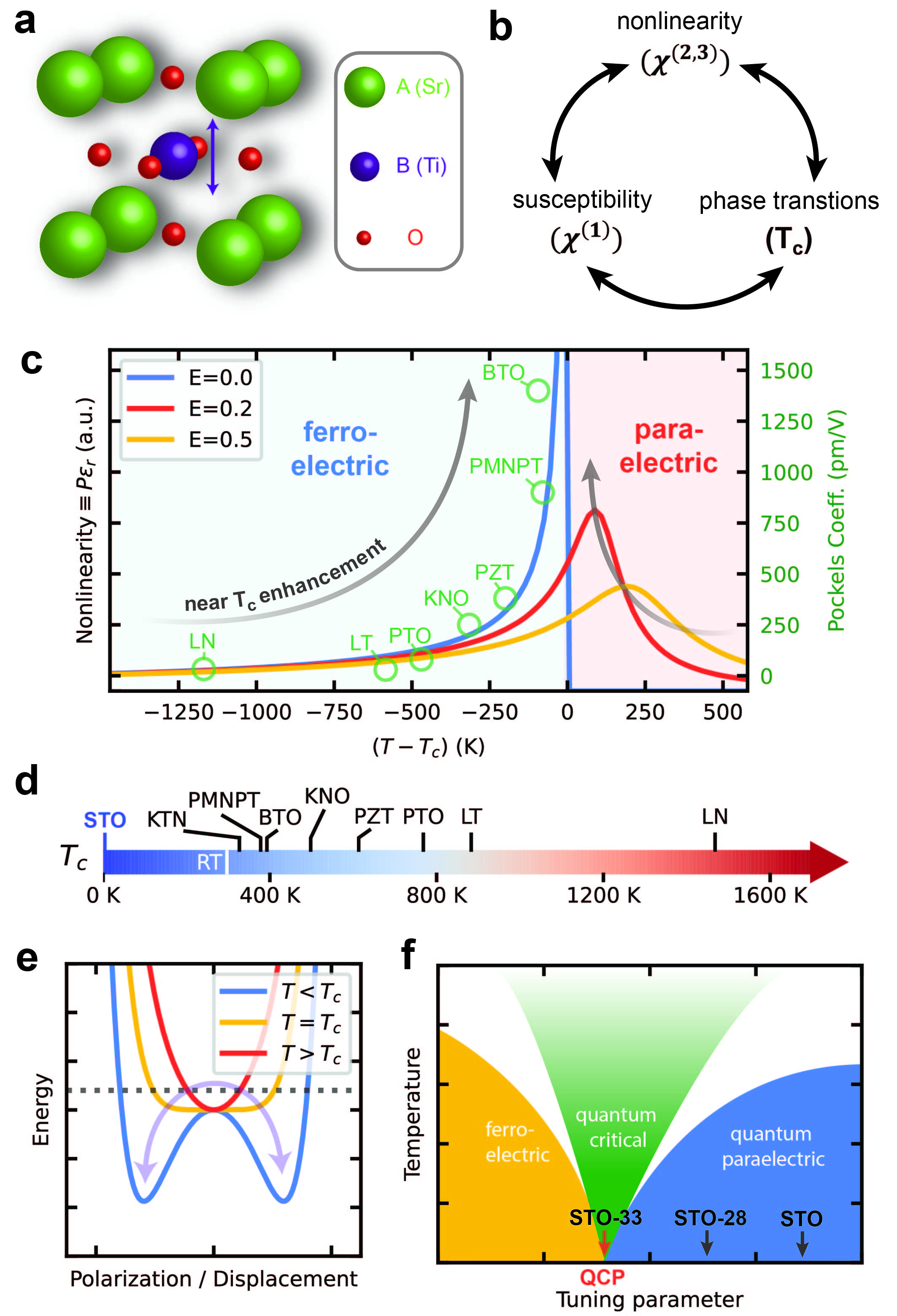}

\caption{\textbf{Searching for nonlinear materials: The link between susceptibility and phase transitions $\vert$} \textbf{a}, Schematic structure of ABO$_3$ perovskites. The motion of the B ($Ti$) ion (purple arrow) mediates the electrical polarization $P$. \textbf{b}, Schematic of the interrelated nature of nonlinearity ($\chi^{(2)}$,$\chi^{(3)}$), susceptibility ($\chi^{(1)}$), and phase transitions at Curie temperature $T_c$. \textbf{c}, Calculated magnitude of $P\epsilon_{r}$ from the Landau-Ginzburg-Devonshire (LGD) model as a function of temperature $T$ and applied field, $E$ (a.u). With $T>T_{c}$ the system is in the paraelectric state, while for $T<T_{c}$ it is ferroelectric. The green circles denote the room temperature Pockels coefficient $r$ for leading ferroelectric perovskites. Approaching $T_c$ from either side of the phase transition results in the largest nonlinearity. \textbf{d}, Comparison of the (bulk) Curie temperature ($T_c$) of different ferroelectric and paraelectric leading electro-optic and piezo-electric materials. \textbf{e}, The Gibbs free energy from the LGD model as a function of the B ion displacement/electrical polarization. In STO, the zero point energy is high enough (horizontal dotted line) that the system never polarizes to one of the two energy wells, shown in blue. Instead, quantum fluctuations cause a stabilization of a nonpolar state (purple arrow). \textbf{f}, Schematic phase diagram for STO. QCP is the quantum critical point. The tuning parameter can arise from multiple mechanisms ($^{18}$O isotope fraction in this work).
}
\label{fig:fig1}
\end{center}
\vspace{0 cm}
\end{figure}

Developments in nonlinear optical materials underpin a wide range of scientific and technical advances, from driving communications worldwide to the creation of entangled photons \cite{Dutt2024,Sinatkas2021}. On the other hand, piezo-electric materials find use in energy harvesting, actuators, medical devices, and beyond \cite{uchino2017advanced}. To date, the leading materials for both of these applications are largely ferroelectrics, commonly of the perovskite structure ABO$_{3}$ (Fig. \ref{fig:fig1}a). While this class of materials share a crystal structure, different materials, such as lithium niobate (LN), lithium tantalate (LT), barium titanate (BTO) and lead zirconate titanate (PZT) have drastically different strengths of these nonlinearities\cite{10.1063/5.0192018,lithiumtantalate,Eltes2020,Wang2018,PZT}. In fact, the strongest nonlinear materials only show a narrow temperature range where the piezo-electric and electro-optic effects are maximized. This is a critical problem for cryogenic and quantum applications, which commonly desire operation temperatures at or below $T=5$ K to match cryostat technology based on liquid helium. 

For example, quantum transducers between the microwave and optical domains require strong nonlinearity in both the optical and/or mechanical domain to create scalable modular quantum computers\cite{linranfan,Jiang2020}. With high nonlinearity, such systems would also enable high speed, low power interconnects for low-temperature applications in high energy physics and superconducting classical computing-- mitigating the wiring and heating problems with microwave cryogenic control and readout \cite{Youssefi2021,Shen2024}. Optimized cryogenic electro-optic optical switches would also form the backbone of photonic-based quantum computing\cite{psiquantum,xanadunature} and provide the needed cryogenic photonic controls for defect centers, trapped atoms, and ions \cite{Mehta2020,coldatoms,photonicinc}. For mechanical systems, strong cryogenic piezo-electric effects are also desired in space-based missions for adaptive optics, actuators, and cryogen fuel transfer, for example\cite{zhang2016cryogenic}. 

Unfortunately, ferroelectric nonlinearities, plasma dispersion and thermo-optic tuning mechanisms usually decrease dramatically at cryogenic temperatures\cite{Eltes2020,Pintus2022}. Therefore, there is a distinct need for nonlinear materials optimized for low temperatures.

\vspace{-0.4cm}
\subsection*{Linking nonlinearity and phase transitions}
\vspace{-0.4cm}

We begin with a theoretical understanding of the origin of both piezo-electric and electro-optic nonlinearities in paraelectric and ferroelectric systems, and use this understanding to guide and optimize the identification of new materials for cryogenic operation (Fig. \ref{fig:fig1}b). 
To start, Miller's rule in optics is an empirical relationship linking nonlinear and linear optical susceptibilities, applicable to a wide range of materials\cite{1074055,physrevb.51.6928,Boydtextbook}. It can be directly linked to a classical Lorentz model under the assumption of weak anharmonic perturbation\cite{Meyer_2024}. Miller's rules state that the nonlinear susceptibilities (such as $\chi^{(2)}$, three-wave mixing) are proportional to the product of the linear susceptibilities ($\chi^{(1)}$) at the frequencies that describe the nonlinearity. Here, the proportionality constant, $\delta$, is related to the electrical polarization, $P$. As a result of Miller's rules, for the electro-optic effect, we find (details in SI Section \ref{sec:SIMiller}):

\vspace{-0.4 cm}
\begin{equation}
    \chi^{(2)}_{EO} \propto n^4 P\epsilon_r
    \label{eq1}
\end{equation}

$n$, $\epsilon_r$ are the index of refraction and quasi-static relative dielectric constant, respectively. With the electro-optic Pockels coefficient $r=2\chi^{(2)}_{EO}/n^4$, then, $r\propto P\epsilon_r$ for most optical materials. Beyond this heuristic, there is an even deeper link between the dielectric behavior and the nonlinearity of paraelectric and ferroelectric perovskites. Surprisingly, the linear electro-optic effect and the linear piezo-electric effect can be directly linked to the corresponding quadratic nonlinearities: the Kerr\cite{Fujiicoeff,10.1063/1.1657458} and electrostrictive effects\cite{Damjanovic_1998, Newnham}. These have a common form across materials. For example, electrostriction is the basis of electromechanical coupling in all insulators \cite{Newnham}. Valid for both the paraelectric and ferroelectric state, the response of either an electrically biased system or a system with a spontaneous or remnant polarization can be framed as the result of a biasing polarization ($P_b$). As a result, we find the Pockels coefficient $r$ and the piezo-electric coefficient $d$, in Voigt notation (SI Section \ref{sec:SIbiasednon}): 

\vspace{-0.4 cm}
\begin{equation}
    r_{ij} \approx 2g_{ij}\epsilon_{r}P_{b,j}
    \label{eq2}
\end{equation}
\vspace{-0.7 cm}
\begin{equation}
    d_{ij} \approx 2q_{ij}\epsilon_{r}P_{b,j}
    \label{eq3}
\end{equation}

The result is that both Pockels and piezo-electric nonlinearities are proportional to the product of the dielectric constant and the bias polarization, the same as the treatment by Miller for optics (Equation \ref{eq1}). We therefore propose this new, general metric for piezo-electricity and electro-optic effects: $P\epsilon_r$. Surprisingly, both Miller's $\delta$ and the coefficients $g$, $q$ have been found to remain within approximately a factor of two for all materials \cite{Boydtextbook,uchino}. This is also why the best piezo-electric materials are often the best electro-optic materials. As most ferroelectrics and paraelectrics have similar maximum polarization, \textit{the main determination of nonlinearity in ferroelectric materials is the dielectric constant}.

%%%%%%%%%%%%%%%%%%%%%%%%%%%%%%%%%%%%%%%%%%%%%%%%%%%%%%%
\begin{figure*}[htb!] 
\begin{center}
\includegraphics[width=2\columnwidth]{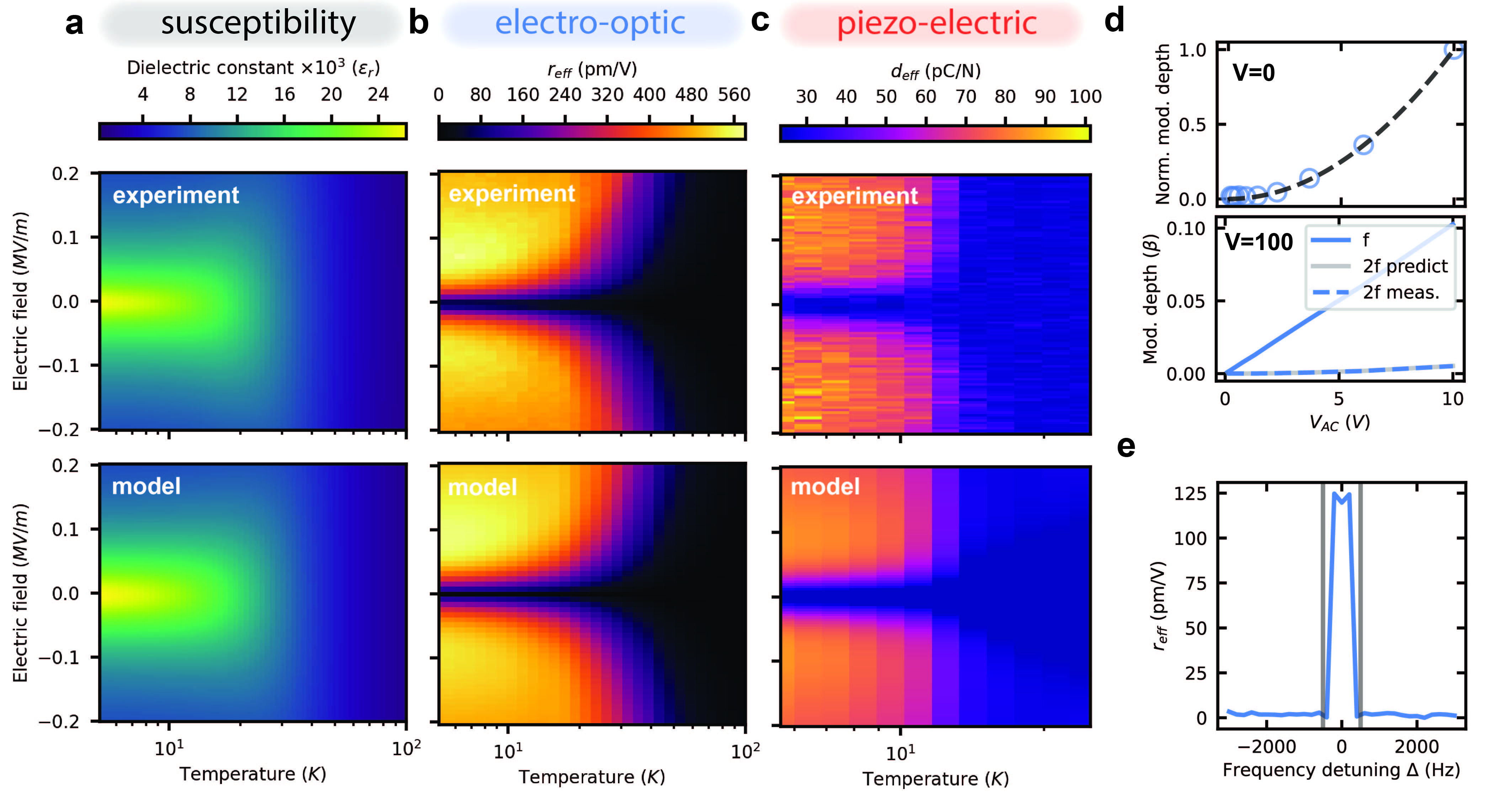}
\vspace{-0.5cm}

\caption{\textbf{Enhanced electro-optic and piezo-electric nonlinearities in natural SrTiO$_3$ $\vert$} \textbf{a}, Data and fitted model of the dielectric constant (susceptibility) as a function of electric field and temperature. \textbf{b}, Using the measured dielectric constant in (\textbf{a}), The experimental Pockels (electro-optic) coefficient as a function of electric field and temperature is fit with one free parameter ($g_{11}$) in the model. \textbf{c}, Using the measured dielectric constant in (\textbf{a}), the experimental piezo-electric coefficient as a function of electric field and temperature is fit with one free parameter ($q_{11}$) in the model. \textbf{d}, At zero bias, the observed electro-optic modulation at twice the modulating frequency ($2f$) (blue dots) is purely quadratic as a function of the modulating voltage (top). With the application of 100 $V$ (0.1 MV/m), the second harmonic is suppressed ($2f$) and the induced modulation observed at $f$ is linear with the modulating voltage, $V_{AC}$ (bottom). \textbf{e}, Effective Pockels coefficient for a weak probe tone when driven with a 100 kHz bias tone at 10 V with detuning $\Delta$. The vertical gray lines show the frequency bandwidth (VBW) of our measurement.  
}
\label{fig:fig2}
\end{center}
\vspace{-0.7cm}
\end{figure*}
%%%%%%%%%%%%%%%%%%%%%%%%%%%%%%%%%%%%%%%%%%%%%%%%%%%%%%

$P \epsilon_r$ can be computed directly as a function of temperature and electric field using a simplified Landau-Ginzburg-Devonshire (LGD) model for the ferroelectric-paraelectric phase transition at Curie temperature $T_c$ (Methods). This results in a deep link between nonlinearity and phase transitions (Fig. \ref{fig:fig1}b,c).  At zero bias, we see in Fig. \ref{fig:fig1}c that no Pockels or piezo-electric nonlinearity will exist above $T_c$ in the nonpolar paraelectric state, as expected. Below $T_c$, we observe a sharp diverging structure that decreases as $T$ becomes lower than $T_c$. Surprisingly, this simple model can fairly well predict the expected nonlinearity of leading electro-optic crystals, shown in Fig. \ref{fig:fig1}c and d. Materials that operate near $T_c$ have enhanced nonlinearity, largely due to a greatly increased dielectric constant, related to the Curie-Weiss law. Intuitively, the increase in dielectric constant near the phase transition can be understood by considering a plot of the Gibbs free energy as a function of electrical polarization (Fig. \ref{fig:fig1}e). As the temperature approaches $T_c$ from above, the energy curve flattens around zero polarization, indicating a low energy cost for a large change in polarization: in other words, an increase in linear susceptibility and dielectric constant. Below $T_c$, the dielectric constant decreases again as the system relaxes into one of two (non-zero) polarization states at the minimum of the Gibbs free energy.

On the other hand, materials in the paraelectric state but still near $T_c$ display large quadratic and induced nonlinearity with an electric field. This result is consistent with reports of giant electro-optic Kerr and electrostrictive effects of paraelectric materials near $T_c$ at room temperature\cite{YU2015429,HIRSHFELD2021104059}. In fact, upon biasing, in Fig. \ref{fig:fig1}c we see that the maximum performance can be pushed into the paraelectric state.

This simplified treatment solidifies the connection of nonlinearity to phase transitions, providing a framework for the discovery of new nonlinear materials. Specifically, nonlinear materials should be chosen largely by the operating temperature of the application. The strongest room temperature piezo-electric and electro-optic active crystals have a $T_c$ near room temperature by design (Fig. \ref{fig:fig1}d). However, no existing materials discovered and optimized for room temperature will be optimal for cryogenic applications. Instead, materials with $T\approx0$ K phase transitions are desired.

\vspace{-0.4cm}
\subsection*{Optical and mechanical nonlinearity in SrTiO$_3$}
\vspace{-0.4cm}

The perovskite strontium titanate (SrTiO$_3$, STO) is exactly such a material (Fig. \ref{fig:fig1}a). STO displays a high index (2.284 at 1550 nm), wide transparency between 400 nm and 2000 nm owing to a wide bandgap (3.25 eV) and is thus suitable for photonics applications\cite{landolt}. In addition, the low temperature phase transition of STO is unique. As STO cools, it never undergoes a paraelectric-to-ferroelectric phase transition, entering instead a new state of ``quantum paraelectricity" where quantum fluctuations suppress the crystallization of a ferroelectric state\cite{MullerPRB,xianli} (Fig. \ref{fig:fig1}e). This behavior is intrinsically linked to the existence of a quantum critical point at zero temperature (Fig. \ref{fig:fig1}f). As a result, it has been subject of extensive studies in condensed matter physics, displaying superconductivity, magnetoelectricity, quantum phase transitions, and floquet driven transient phases\cite{Kleemann2020,Rischau2017,xianli}. 

As STO cools, its low-frequency dielectric constant rises, and becomes one of the largest reported for any crystal at any temperature at over 25,000 (Fig. \ref{fig:fig2}a). However, its quantum paraelectric nature causes a saturation of the dielectric response at 10 K and lower, deviating from the traditional Curie-Weiss law. Therefore, this material on the verge of ferroelectric instability results in a massive but stable electrical susceptibility. In this work, we probe the link of this unique quantum phase to optical and mechanical nonlinearities for the first time.

We first plot the bias field and temperature dependence of the dielectric response of natural STO (Fig. \ref{fig:fig2}a). Using a modified LGD model with a quantum paraelectric correction following Zubko and Vendik\cite{Vendik1997,Vendik1999} (Methods and SI Section \ref{sec:SIpol}), we are able to show remarkable agreement with the observed data. From the measured dielectric response as a function of bias field, we calculate the polarization, and therefore $P\epsilon_r$, directly from our experimental data.

%%%%%%
\begin{figure*}[htb!] 
\begin{center}
\includegraphics[width=2\columnwidth]{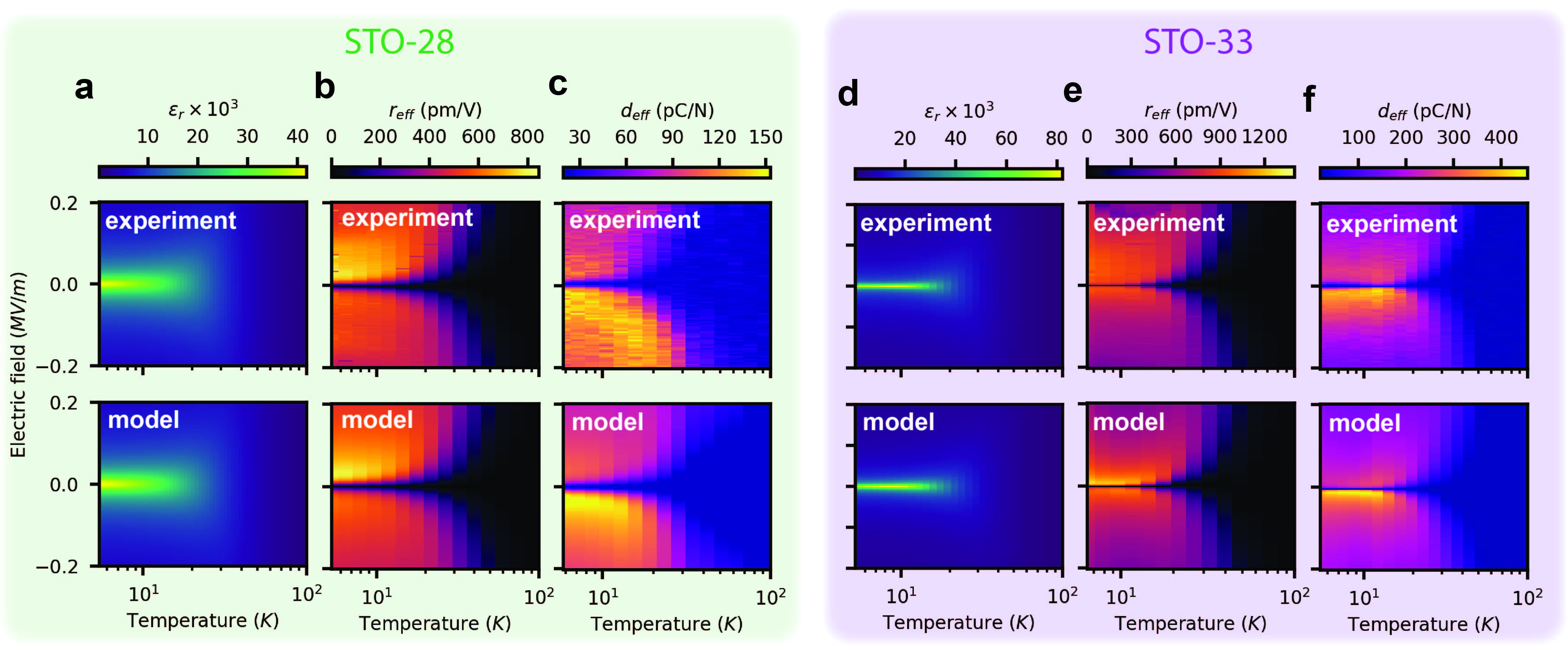}
\caption{\textbf{Increasing optical and piezo-electric nonlinearities by tuning towards quantum criticality $\vert$} \textbf{a}, Data and fitted model of the dielectric constant (susceptibility) as a function of electric field and temperature for STO-28. \textbf{b}, Using the measured dielectric constant in (\textbf{a}), the experimental Pockels coefficient as a function of electric field and temperature is fit with one free parameter, $g_{11}$. \textbf{c}, Using the measured dielectric constant in (\textbf{a}), the experimental piezo-electric coefficient as a function of electric field and temperature is fit with one free parameter, $q_{11}$. \textbf{d},\textbf{e},\textbf{f} are the same as \textbf{a},\textbf{b},\textbf{c} but for STO-33.
}
\label{fig:fig3}
\end{center}
\vspace{-0.6cm}
\end{figure*}
%%%%%%%%%%%%%%%%%%%%%%%%%%%%%%%%%%%%%%%%%%%%%%%%%%%%%%

We then measure the effective Pockels nonlinearity of STO using a cryogenic Mach-Zehnder-Interferometer (MZI), which extracts the phase shift from the index modulation of the STO crystal as a function of electrical bias (Methods and SI Section \ref{sec:SIMZI}). The measured Pockels coefficient ($r_{33}$) induced by the applied bias to the sample is shown in Fig. \ref{fig:fig2}b. We observe $r>500$ pm/V for a wide range of voltages at cryogenic temperatures. This is the largest measured Pockels coefficient at these temperatures, outperforming cryogenic BTO (170 pm/V)\cite{Eltes2020} and electro-optic polymers (150 pm/V)\cite{Park:10}. In addition, it is more than a factor of 20 greater than commonly used LN (24 pm/V)\cite{LN_lowtemp}. 

The observed strength and shape of our observed data are fit with a single free parameter, $g_{11}$, and the measured $P\epsilon_r$ extracted from Fig. \ref{fig:fig2}a. There is a remarkable match with our model, once again verifying the link between dielectric and nonlinear optical properties. We fit $g_{11}=0.15(2)$ at 5 K, which is close to the previously reported value at room temperature of $g_{11}=0.14$ \cite{Fujiicoeff}. We note that the required field to bias to maximum nonlinearity is only 0.1 MV/m, such that 1 V across 10 $\mu$m is more than sufficient for biased operation in integrated waveguides.

Next, we focus on the bias-induced linear (converse) piezo-electric effect in STO utilizing the same MZI in a modified configuration (SI Section \ref{sec:SIMZI} and Fig.~\ref{fig:s4_exp}b) to infer the phase shift from the moving boundary of the STO as function of bias (Fig. \ref{fig:fig2}c). We measure $d_{33}$= 90(6) pC/N, which is over an order of magnitude greater than LN, and exceeds all known piezo-electric coefficients at low temperature. Similarly, this response is very well modeled through the observed $P\epsilon_r$ and fitting $q_{11}=0.024(3)$ at 5 K, which falls within the range of related ferroelectric materials \cite{10.1063/1.4861260}. Crucially, the electrostrictive nature of STO may result in advantages over traditional piezo-electrics, thanks to low hysteresis, low creep or aging effects, fast response, and no traditional poling requirement\cite{uchino}.

Our extracted $d_{33}$ value is also consistent with density functional theory predictions of ferroelectric STO\cite{PhysRevB.88.035102}. We note that other previous work has reported extremely large effective cryogenic piezo-electric coefficients on electrically biased STO of 1600 $pC/N$\cite{grupp}. However, follow-up measurements revealed calibration issues with a contact strain sensor, which caused an incorrect estimation of the piezoelectric coefficient\cite{Grupp1999}. Our measurements are instead interferometric (independent of detector response) and non-contact, bringing clarity to this longstanding result.

%%%%%%%%%%%%%%%%%%%%%%%%%%%%%%%%%%%%%%%%%%%%%%%%%
Our data is taken at a lower frequency (100 kHz) than many applications require, which are typically in the MHz to GHz regime. This is to avoid a RC cutoff at about 1 MHz, entirely limited by the sample capacitance of our large device (SI Section \ref{sec:SIfreq}). As the dielectric response remains strong to GHz frequencies \cite{Neville,Geyer,imec}, the nonlinearity should still be similarly strong. Using the Lyddane-Sachs-Teller relation, we predict a fundamental electro-optic bandwidth above 1 THz for all samples (SI Section \ref{sec:SIEOBand} and Fig. \ref{fig:s15}). This indicates that the strong, fast, electro-optic response can be preserved in properly designed nanophotonic devices with reduced dimensions.   

Without a DC bias field, STO is a quadratic electro-optic material, with a large $\chi^{(3)}(0,0,\omega,\omega)$, where $\omega$ is an optical frequency (SI Section \ref{sec:SIquadindex}). Therefore, a quasi-static electrical modulation at $f$ can produce optical phase modulation at $2f$, a form of frequency doubling. We measure this second-harmonic modulation at zero bias as a function of drive voltage, shown in Fig. \ref{fig:fig2}d. The strength of the modulation is quadratic in voltage, as expected. In Fig. \ref{fig:fig2}d, we also show that when an electrical bias is applied, the modulation at $f$ from the linear electro-optic effect becomes dominant, and the residual $2f$ modulation arises purely from higher harmonics of the strong drive. The quadratic electro optic $s_{ij}$ and biased linear $r_{ij}$ coefficient can be related by $r_{ij}=2 s_{ij}E_{b,j}$. From this, we calculate $s_{11}$ to be at least 5$\times$10$^{-15}$ m$^2$/V$^2$ or higher for natural STO. This is on par or higher than even the best potassium tantalate niobate (KTN) quadratic EO devices\cite{Fujiura}, which display the largest EO Kerr effect found among all known materials.

For traditional quadratic electro-optic materials such as KTN, optimal performance requires stabilization of the composition on the order of 1\%, and an optimal operation temperature range only a few $\degree$C wide \cite{Fujiura,WANG2023838}. In contrast, because of the quantum paraelectric phase, the electro-optic response of STO is stable over a wide range of temperatures and requires no compositional tuning. These results may also be extended to higher temperature in strained films, which have already been tuned to ferroelectricity at ambient temperatures\cite{Haeni2004}. In fact, STO has a predicted Pockels elements as large as 500 pm/V if strained \cite{PhysRevMaterials.2.115202}, a close match to the observed data here.

Because the linear Pockels effect in quantum paraelectric STO is induced via a bias field, there are opportunities to dynamically tune the nonlinearity. For example, in Fig. \ref{fig:fig2}f we demonstrate strong modulation occurs only with the signal frequency is matched with the modulation frequency. This capability can be engineered spatially and in time, enabling dynamic periodic poling and bipolar temporal biasing which can inhibit photorefractive optical damage and charge buildup\cite{HIRSHFELD2021104059}. Finally, STO also avoids some of the difficulties with ferroelectric BTO. For example, STO is paraelectric, so there is no multi-domain complexity or potential scattering loss from domain boundaries (SI Section \ref{sec:SILoss}). Moreover, it is not necessary to orient an optical device with respect to a specific crystal axis in STO. These geometric considerations, combined with the ability to use the same electrodes to apply both the signal tone and ``poling" field, ensure a maximal utilization of the full strength of the Pockels tensor element, a difficulty with BTO. 

\vspace{-0.4cm}
\subsection*{Enhancing nonlinearity near quantum criticality}
\vspace{-0.4cm}

Previous works have shown that substituting $^{18}$O for naturally occurring $^{16}$O tunes the system towards and through quantum criticality (Fig. \ref{fig:fig1}f)\cite{itoh,Rowley2014}. At about 33\% concentration of $^{18}$O, the quantum critical point (QCP) is reached where the dielectric properties are further enhanced near $T=0$, no longer suppressed by quantum fluctuations. We therefore hypothesize that the cryogenic piezo-electric and electro-optic nonlinearities will grow as well, and study them for the first time.

We first measure the dielectric properties of SrTi$(^{16}O_{0.72}{^{18}O_{0.28}})_3$ (``STO-28") and SrTi$(^{16}O_{0.67}{^{18}O_{0.33}})_3$ (``STO-33"), prepared by annealing natural STO (``STO") in isotopically purified oxygen gas (Methods), shown in Fig. \ref{fig:fig3}a,d. The same modified LGD model for natural STO fits the response in these samples. We notice a sharper dependence of the dielectric constant with respect to the electric field, with a corresponding higher peak value, of around 42,000 and 82,000 for STO-28 and STO-33, respectively (see Table \ref{tab:sample_parameters}).  Interestingly, despite the large zero-bias dielectric constant, the achievable electrical polarization is quite similar at about 2 $\mu$C/cm$^{2}$ among all samples (Fig. \ref{fig:s1}). 

We perform measurements of the bias-induced electro-optic and piezo-electric nonlinearities in the same way as natural STO, shown in Fig. \ref{fig:fig3}, alongside the model from the measured dielectric data. Once again, there is a remarkable match between the intuition from the ``$P\epsilon_r$" rule and the link between quadratic and linear nonlinearities. This includes a replication of the slight asymmetry, which is also present in the dielectric data. For STO-28, we report maximal values of $r_{33}=840(50)$ and $d_{33}=150(9)$ pC/N. For STO-33, we measure $r_{33}=1150(30)$ pm/V and $d_{33}=435(25)$ pC/N at their peak (SI Section \ref{sec:SISTO33}). As STO approaches the QCP, we therefore observe an almost doubling of the Pockels effect and over a four-time improvement in the piezo-electric effect, approaching the values of even the strongest room temperature electro-optic and piezo-electric crystals. We also show a line cut of the $T=5$ K dependence of the the nonlinearity with applied bias for natural and isotope-exchanged STO in Fig. \ref{fig:fig4}a. The required bias fields for maximum nonlinearity decrease at the system is tuned to criticality, while surprisingly, the ``high bias" electro-optic coefficient is almost identical between the samples.

We also report the electromechanical coupling factor, $k_{33}$ among all samples (SI Section \ref{sec:SIelectromech} and Fig. \ref{fig:s2}), which more accurately describes the figure of merit for converting electrical energy to mechanical energy at cryogenic temperatures, $k_{33}$ roughly triples between STO and STO-33, and is only limited by the strength of the bias field.

Finally, we utilize the large quadratic electro-optic effect in isotopically exchanged STO to make a highly efficient cryogenic beam steering device that deflects a laser by 10 mrad at a field of 0.2 MV/m, (Fig. \ref{fig:s5} and SI Section \ref{sec:SIsteering}). This is on par with the best room temperature KTN beam deflectors, and could find use in space communications or efficient cryogenic varifocal lenses \cite{Imai2009}. 

\begin{table}[ht!]
\vspace{-0.25cm}
\caption{\label{tab:sample_parameters} \textbf{Summary of parameters for natural and isotopically engineered SrTiO$_3$ $\vert$} $E_{max}$ is the electric field where $r_{33}$ is maximized. All parameters are at $T=5$ K or the temperature where maximized ($T<10$ K for all). }
\begin{ruledtabular}
\begin{tabular}{lccc}
\textbf{} & \textbf{STO} & \textbf{STO-28} & \textbf{STO-33} \\ 
\hline 
$r_{33,max}$ (pm/V)       & 580(60)            & 840(50)             & 1150(30)              \\
$d_{33,max}$ (pC/N)       & 91(6)             & 150(9)              & 435(25)            \\
$\epsilon_{r,max}$        & 26.0(1.5) k             & 42.0(3.7) k           & 82.0(1.6) k           \\
$(P \epsilon_r)_{max}$ (C/m$^2$)       & 200(20)            & 145(20)            & 164(5)            \\
$E_{max}$ (MV/m)        & 0.080              & 0.036              & 0.008              \\
$g_{11}$ (m$^4$/C$^2$)       & 0.15(2)            & 0.21(3)            & 0.30(1)           \\
$q_{11}$ (m$^4$/C$^2$)       & 0.024(3)            & 0.039(6)              & 0.110(7)            \\
$k_{33,max}$       & 0.15(1)              & 0.21(2)              & 0.46(3)             \\

\end{tabular}
\end{ruledtabular}
\end{table}

\vspace{-0.4cm}
\subsection*{Trends in nonlinear materials}
\vspace{-0.4cm}

\begin{figure}[htb!] 
\begin{center}
\includegraphics[width=0.985\columnwidth]{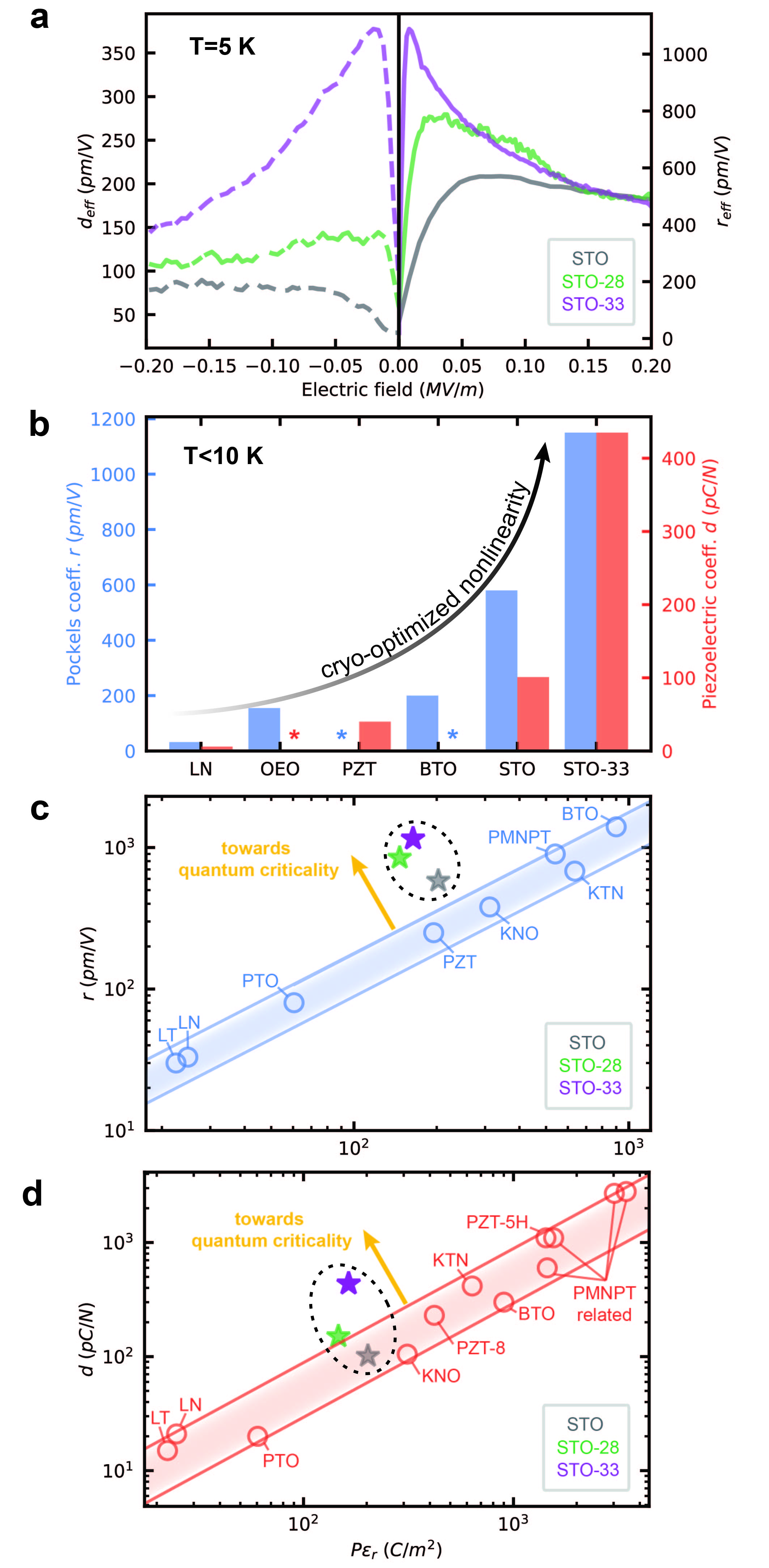}
\vspace{-0.2cm}
\caption{\textbf{Boosted nonlinearities in SrTiO$_3$ and comparisons to existing materials $\vert$} \textbf{a}, Field dependent Pockels (solid, right) and piezo-electric coefficients (dashed, left). \textbf{b}, Cryogenic ($T<10$ K) electro-optic and piezoelectric coefficients of leading materials. ``*" denote that the value is unknown. \textbf{c}, Scaling of the Pockels effect with $P\epsilon_r$ for a range of ferroelectric materials (blue circles). The blue shaded region is a linear scaling within $0.05<g<0.10$ \textbf{d}, Scaling of the piezo-electric effect with $P\epsilon_r$ for a range of ferroelectric materials (red circles). The red shaded region is a linear scaling within $0.01<q<0.04$. All values are reported at room temperature, except for STO. References for the reported values are found in the Supplemental information.
}
\label{fig:fig4}
\end{center}
\vspace{-0.2cm}
\end{figure}

To better understand these results in the landscape of nonlinear materials, we compare the performance of natural and isotopically engineered STO with other piezo-electric and Pockels materials (more details in SI Section \ref{sec:SIcomparison}). First, we compare our results to existing reports of cryogenic ($T<10$ K) nonlinearities in Fig. \ref{fig:fig4}b, showing at least a 5$\times$ improvement over all previously reported materials, and more than a 40-fold improvement over lithium niobate. We then show the measured $r$ and $d$ as a function of $P\epsilon_r$ for a wide range of perovskite materials measured \textit{at room temperature} (Figures \ref{fig:fig4}c and \ref{fig:fig4}d). Following the insights outlined in this work, indeed these materials do fall within a narrow band, linearly dependent on the known $P\epsilon_r$ and reflecting the similar $g$, $q$ among materials. 

%%%%%%%%%%%%%%%%%%%%%%%%%%%%%%%%%%%%%%%%%%%%%%%%%%%%%%%
%%%%%%%%%%%%%%%%%%%%%%%%%%%%%%%%%%%%%%%%%%%%%%%%%%%%%%

Therefore, up to a factor of order unity, $P\epsilon_{r}$ is predictive of the strength of the electro-optic and piezo-electric response. Other works have suggested the scaling of piezo-electric materials with dielectric constant, but this relies on the similar electrical polarization found in ferroelectrics. By instead framing as a dependency on $P\epsilon_r$, not only is the trend in materials better described, but biased paraelectric materials can be directly compared. The notable exception to these trends is STO, which lies outside the band for the electro-optic effect in natural STO and further deviates as it is tuned to quantum criticality. For piezo-electricity, STO starts within the band of known materials, but is tuned outside the range at criticality. Despite the large changes in the peak dielectric constant, the maximum value of $P\epsilon_r$ remains very similar among all our samples. Therefore, the increased nonlinearity arises from an unexpected and significant change in $g$ and $q$, which are nominally similar for most materials (Table \ref{tab:sample_parameters} and SI Section \ref{sec:SItrendgq}). 

In traditional ferroelectric-paraelectric materials, the product of the Curie-Weiss constant, and the electrostrictive constant $q$, have been shown to be almost the same over all materials \cite{uchino}. In STO, the connection is more intricate, as well known scaling laws do not hold. For example, even the mechanism for tuning the phase transition in STO by stabilizing quantum fluctuations differs fundamentally from a compositionally tuned ferroelectric materials like PZT or KTN \cite{wangsto}. Therefore, we observe a distinct difference in quantum critical STO (near a quantum phase transition) compared to other perovskites which may be close to a classical phase transition (Fig. \ref{fig:fig4}c,d). 

As a result, the measured electrostrictive response for the isotopically substituted STO can also be classified as ``giant"\cite{giantQ}, exceeding the heuristic scaling of known materials with the dielectric constant and the elastic compliance by more than one order of magnitude, and is therefore a ``non-Newnham" electrostrictive material\cite{Newnham}.

\vspace{-0.4cm}
\subsection*{Discussion}
\vspace{-0.4cm}

%%%
In this work, we propose a universal metric to understand second order optical and mechanical nonlinearities in materials, $P\epsilon_r$, whose change describes the full temperature and electrical bias dependence of a material's nonlinearity-- and is enhanced near phase transitions. Guided by this insight, we identify STO as an ideal material for cryogenic applications. We demonstrate large piezo-electric and electro-optic nonlinearities in STO that are further enhanced by tuning to quantum criticality. Surprisingly, while the observed behavior follows predictions, the scaling tensor coefficients $g$ and $q$ become drastically different from traditional ferroelectrics as STO approaches the QCP. Our measurements are also free from potential strain and clamping effects which may be present in grown films, and constitute the largest nonlinearities measured at cryogenic temperatures.

The observed strong nonlinearities unlock a wide range of opportunities for cryogenic classical and quantum applications. For example, the large Pockels effect reported here enables ultra-low V$_\pi$L modulators for cryogenic photonics and efficient interconnects for superconducting microwave circuits\cite{Shen2024}. Such a highly efficient modulator would allow for smaller device sizes with corresponding exponentially less loss and lower power consumption, critical to unlock the large-scale complex cryogenic photonic circuits desired for photonic quantum computing, for example\cite{moodyroad2022, xanadunature} (additional discussion in SI Section \ref{sec:SIoutlook}). On the other hand, cryogenic piezo-electric systems could find use in space missions, low temperature scanners, cryofluid sensors and actuators\cite{zhang2016cryogenic}. Importantly, STO is foundry-compatible and lead-free, eliminating many of the processing and safety issues related to lithium niobate and lead-containing piezo-electrics.

Compared to traditional ferroelectric materials, quantum paraelectric materials biased to maximal nonlinearity have a much lower polarization and higher dielectric constant. An alternate pathway to nonlinearity would be to tune fully into the ferroelectric phase, where STO would display similarly large polarization while maintaining enhanced cryogenic nonlinearities. While our measurements are on bulk crystals of STO, scaling and integration with silicon electronic and photonic circuits is also feasible, with high quality Molecular Beam Epitaxy (MBE) at the wafer scale already demonstrated \cite{Brooks}. Very recent reports have also measured ferroelectric STO thin films that outperform cryogenic BTO \cite{imec}.

Our work highlights the advantages to exploring new photonic and piezo-electric materials that utilize exotic condensed matter physics to remove roadblocks for cryogenic technologies. It also suggests the exploration of related quantum paraelectric materials and low $T_c$ ferroelectrics for improved cryogenic optical and mechanical devices.

\vspace{-0.4cm}
\subsection*{Methods}

\vspace{-0.4cm}
\subsubsection*{Landau-Ginzburg-Devonshire model}
\vspace{-0.4cm}

In general, we can understand the enhancement of nonlinearity near a phase transition by examining the Landau-Ginzburg-Devonshire phenomenological theory, which has been found to well describe the critical phenomena in ferroelectric-paraelectric systems\cite{Damjanovic_1998,LIU2023135}. Here, the polarization ($P$) and electric field ($E_b$ dependent Gibbs free energy, $G$ is:

\begin{equation}
    G= \frac{1}{2} \alpha P^2+\frac{1}{4} \beta P^4+\frac{1}{6} \gamma P^6-E_b P
    \label{Gibbs}
\end{equation}

The equilibrium state is found where the Gibbs free energy is minimized, and the dielectric constant is found through:

\begin{equation}
   \frac{1}{(\epsilon_r-1)\epsilon_0}=\frac{\partial^2 G}{\partial P^2}
    \label{Gibbs dielectric}
\end{equation}

Usually, $\beta$ and $\gamma$ are temperature independent or weakly dependent on temperature, while $\alpha\approx\alpha_0 (T-T_c)$ describes the effect of temperature. If $\beta$ negative then the transition is considered first order, if positive then second order. In our simulations, we set $\alpha_0=1$, $\beta=0.01$, $\gamma=1$.

\vspace{-0.4cm}
\subsubsection*{Zubko and Vendik extension for quantum paraelectrics}
\vspace{-0.4cm}

A variety of models have been used to describe the bias and temperature dependent response of the dielectric constant of STO. The best match with experiment\cite{Fujishita2016} comes from a solution of the Landau-Ginzburg-Devonshire equation by Vendik and Zubko\cite{Vendik1997,10.1063/1.366180}, resulting in:

\begin{equation}
\epsilon = \frac{C/T_c}{\left[\sqrt{\xi^2 + \eta^3 + \xi} \right]^{2/3} + \left[\sqrt{\xi^2 + \eta^3 - \xi} \right]^{2/3} - \eta}
    \label{Vendik}
\end{equation}

Where $C$ is the effective Curie-Weiss constant, $\xi^2=\xi^2_B+\xi^2_0$ with $\xi_B = E/E_0$, where $E$ is the applied bias and $E_0$ the renormalizing field. $\eta$ is an integral which can be approximated as $\frac{T_D}{T_c} \sqrt{\frac{1}{16}+(\frac{T}{T_D})^2}-1$ with $T_D$ as the Debye temperature of the relevant mode and $T_c$ as the Curie temperature.

In principle, one value of $T_C$, $T_D$, $E_0$, $C$ and $\xi_0$ are needed to fit the entire temperature and voltage dependence of the dielectric constant. We first fit the temperature dependence under zero bias to fit all parameters except $E_0$, which is explicitly related to the electric field dependence (Fig. \ref{fig:s1}). The entire dataset is then globally fitted to extract $E_0$, showing a very close match with the observed data (Fig. \ref{fig:s1}). Our fit also allows for a small temperature-dependent offset field which reflects any built-in polarization (arising from strain or ferroelectricity, for example) which drives the asymmetry in Fig. \ref{fig:fig3}, for example.

\vspace{-0.4cm}
\subsubsection*{Dielectric measurements}
\vspace{-0.4cm}
Dielectric constants are extracted from the capacitance measured with a LCR meter and assume a parallel plate model, $C=\epsilon_0\epsilon_r A/d$, where $A$ is the plate area and $d$ the thickness of the sample. This is a reasonable assumption due to the dimensions of the sample and the high dielectric constant which reduces fringing fields. A background capacitance is subtracted. (100) STO is used for all measurements so that the observed dielectric constants are all $\epsilon_{100}$. With the dielectric response, $dP/dE=(\epsilon_{r}-1)\epsilon_0$, electrical polarization is found with:

\begin{equation}
    P(E_b,T)=\int_{0}^{E_b} (\epsilon(E,T)-1)\epsilon_0 dE
\end{equation}

All measurements are taken in a standard ``P-E" hysteresis loop. All dielectric constant data was taken at 100 kHz with a drive voltage of 1 V. The electro-optic and piezo-electric data was also taken with a hysteresis loop. Reported data is an average over that loop.

\vspace{-0.4cm}
\subsubsection*{Variable temperature interferometric measurement}
\vspace{-0.4cm}

We measure the piezo-electric and electro-optic effect utilizing a sensitive Mach-Zehnder-Interferometer (MZI). Samples are mounted inside of a Montana Instruments S200 cryostat with a base temperature of T = 5 K and with rapid sample temperature control and electrical connections. A schematic is shown in Supplemental information Fig. \ref{fig:s4_exp}. The backside is affixed with a flexible but conductive adhesive, and the top surface is wire-bonded. The samples are metalized on the top and bottom surfaces with 30 nm of Pt and 50 nm of Au. The large work function of Pt avoids charge injection into the STO sample. The samples are commercially available 5x5 mm chips either 0.5 or 1 mm thick. A 1550 nm telecom laser (Anritsu MG9638A) is passed through a fiber beamsplitter. One arm is outcoupled in free space, passes first through a linear polarizer (LP) and is then focused onto the STO with a 200 mm lens. After transmitting through the STO the beam is re-collimated by another 200 mm lens, and passes through another linear polarizer, set to the same direction as the first LP. The other arm passes through a fiber coupled acousto-optic modulator (AOM, model 1550-AOM-2 manufactured by Aerodiode), which shifts the optical frequency by 200 MHz. The two arms are then recombined with a fiber beamsplitter, and the output power of one arm is monitored with a 50 $\ohm$ terminated fast photo-diode (Thorlabs DET08CFC), and sent to a spectrum analyzer (Agilent N9030A). 

Without interacting with the sample, the signal displays a beat note at the AOM frequency of 200 MHz, shifted away from DC noise and background. To measure the electro-optic effect, two side facets are polished to high optical quality and flatness to allow for laser transmission between the electrodes. An oscillating voltage at frequency $f_{mod}$ is applied to the electrodes. The oscillating change of index due to the Pockels effect results in an oscillating phase shift which manifests as a further modulation on the power, creating signal at $f_{AOM}\pm f_{mod}$. By measuring the ratio of the power at $f_{AOM}$ and $f_{AOM}\pm f_{mod}$, a calibration free measurement can be made of the modulation index $\beta$, which characterizes the strength of the phase shift (Supplemental information). 

Measurement of the piezo-electric effect proceeds similarly, but with the laser reflected off the top metalized surface of the sample at a 45$^o$ angle. The moving ``mirror" of the sample due to the applied bias causes an oscillating phase shift that can similarly be picked up by the interferometer. Typical laser powers are on the order of 1 mW at 1550 nm with a beam waist of order 100 $\mu$m. The sample is mounted on a three-axis cryogenic nano-positioner which enables precise centering and alignment of the sample with respect to the beam. Electrical bias to the sample consists of both AC (Rigol DG4102) and DC (Keithley 2400) which are combined with a high voltage bias tee (Picosecond Pulse Labs 5530B). For all samples, at the highest voltages, the current never exceeds 25 nA. The strength of the electro-optic effect is independent of power between 0.1 mW and 6 mW, such that photorefractive screening is negligible. Further details can be found in the Supplementary information.

We calibrate our measurements by using a control chip of z-cut LiNbO$_3$ (LN), measuring $d_{33}=4.6(3)$ pm/V. This is about a 25\% underestimate compared to the known value of 6 pm/V. We attribute this difference to the partially free moving boundary of the adhesive used to affix the sample. Mounting the sample multiple times did not change this underestimate. We therefore report here the piezo-electric coefficients of STO as a \textit{lower bound}. This ``loose" mounting also allows us to report coefficients nominally unaffected by strain. The frequency of piezo-electric measurements ranged from 20 kHz to 100 kHz, depending on the sample, to avoid sharp mechanical resonances. 

The electro-optic nonlinearities do not suffer from this mounting detail. In LN, we measure $r_{33}\approx$ 31 pm/V and $r_{13}\approx$ 9 pm/V. These are within 10\% of the reported values for LN. Our estimated uncertainty in the size of the electrode/length of the beam path is on this order of magnitude. All electro-optic data was taken at 100 kHz with a peak-to-peak drive voltage of 1 V. We observe that the calculated Pockels coefficient does not depend on the drive voltage between 0.2 V to 10 V.

Measurements were corrected for bias-tee and cable losses of the drive signals of about 1.3 dB. Our measurement is based on a power modulation observed at the output of the interferometer. For each experiment, the interferometer is bypassed to observe the sample-induced modulation of the laser intensity. For all experiments, the magnitude of this ``pure power" modulation is at least 30 dB lower than the observed modulation due to the ``true" induced phase shift. We estimate that this effect contributes to an overestimate of the nonlinearity by up to 3\% at maximum.

 Due to the quasi-cubic nature of STO, we report $r$ and $d$ as $r_{33}$ and $d_{33}$ but the corresponding $q$ and $g$ as $g_{11}$ and $q_{11}$, to match convention from other works. We ignore the effects of antiferrodistortive structural phase transitions which occur at around 105 K.

\vspace{-0.4cm}
\subsubsection*{Isotopically exchanged STO}
\vspace{-0.4cm}

Isotope substitution in STO samples is achieved through a high temperature annealing treatment in isotopically enriched oxygen gas. We begin with commercially available bulk STO chips, which are loaded into a furnace that is evacuated to a vacuum level of $~5\times10^{-6}$ Torr with a turbomolecular pump. Then, the sample chamber is filled with 0.5 atm of $^{18}$O gas, and the sample is heated to $1000 ^o$C at a constant rate, for 300 minutes. The sample remains in the $^{18}$O rich environment at $1000 ^o$C for approximately 14 days (19998 minutes). At this higher temperature, the $^{18}$O pressure is $\approx$0.9 atm, and it is kept constant for the entire duration of the process. Subsequently, the sample is cooled down to room temperature for approximately 360 minutes by turning the heater off, and the chamber is pumped to the initial vacuum levels. To ensure an even distribution of the $^{18}$O isotopes throughout the bulk crystal, an additional vacuum annealing step is performed. This step involves heating to $1000 ^o$C in 300 minutes, and maintaining the sample at low vacuum levels and high temperature for 14 days. At the end of the process, the samples weight is recorded, and it is compared to the weight of the sample before the process to get an estimate of the $^{18}$O isotope concentration. Details regarding the total duration of the $^{18}$O and vacuum anneals, as well as the $^{18}$O estimate based on the weight measurement for all samples can be found in Table~\ref{tab:ISO} in the Supplementary information.

Isotope content is verified using nanoscale secondary ion mass spectrometry (nanoSIMS), with the Cameca NanoSIMS 50L at the Stanford Nano Shared Facilities, and cross-checked with the weight-based measurement (Supplemental Information). Analyses are performed using a Cs$^+$ primary ion beam set to an operating current of 5 pA and total impact energy of 16 keV. Detectors are set to measure masses $^{16}$O$^{-}$ and $^{18}$O$^{-}$ at a mass resolution of $>$6000 (Cameca definition) to avoid interferences. Before the measurements, we coat the entire sample with 10nm of gold for conductivity and perform a 20-30 minute pre-sputtering step inside the NanoSIMS 50L to remove superficial gold and etch into the STO. The measurements are performed by raster scanning multiple areas 25$\times$25$~\mu m$ in size, at a resolution of 256$\times$256 pixels per data frame. Each analysis consists of 20-60 frames of data that are stacked to improve precision, and the first 5-20 frames are discarded as the beam sputters through the surface coatings and reaches a steady state of Cs+ implantation. The error on our measurement of the isotope concentration is calculated based on the recorded variance in $^{16}$O and $^{18}$O concentrations and is less than $0.5$\% in all measurements (SI Section
\ref{sec:SI18O}).

\vspace{-0.4cm}
\subsection*{Data availability}
\vspace{-0.4cm}

The experimental data and numerical simulations presented here are available from the corresponding authors upon reasonable request.

\putbib[main] 
\end{bibunit}

\newpage
\subsection*{Acknowledgments} 
\vspace{-0.4cm}

This work was supported by a Vannevar Bush Faculty Fellowship from the US Department of Defense, Samsung Electronics, and by the Department of Energy under the Q-NEXT program and Grant No. DE-SC0020115E. This work was partially funded by an unrestricted gift from Google. C.P.A. acknowledges support by an appointment to the Intelligence Community Postdoctoral Research Fellowship Program at Stanford University administered by Oak Ridge Institute for Science and Education (ORISE) through an interagency agreement between the U.S. Department of Energy and the Office of the Director of National Intelligence (ODNI). G.S. acknowledges support from the Stanford Bloch Postdoctoral Fellowship. S.E. acknowledges support from the Shoucheng Zhang Graduate Fellowship and Korea Foundation for Advanced Studies Overseas Ph.D. Fellowship. Part of this work was performed at the Stanford Nanofabrication Facility (SNF) and the Stanford Nano Shared Facilities (SNSF), supported by the NSF under Grant No. ECCS-2026822. The work at the University of Michigan is supported by the National Science Foundation under Award No.DMR-2317618 (sample isotope annealing) and is supported by the Department of Energy under Award No. DE-SC0020184 (dielectric characterization).

We thank Daniil Lukin, Melissa Guidry, Joshua Yang, Zelong Yin and Yujia Yuan for assistance with characterizing initial SrTiO$_3$ materials. We thank Erik Szakiel, Wentao Jiang, and Sahil Dagli for discussions and fabrication assistance on related devices. We thank Jason (Shangqing) Huang and Eric Rosenthal for comments on the manuscript.\\

\textit{Note added.}--- During preparation of this work, we were made aware of complementary work measuring cryogenic properties of STO thin films \cite{imec}.

\vspace{-0.4cm}
\subsection*{Author contributions}
\vspace{-0.4cm}

C.P.A. and G.S. conceived and designed the experiment. C.P.A. and G.S. built the measurement system, performed the experiment, and analyzed the data. A.C. and L.L. created the isotopically substituted STO. K.V.G. and S.E. assisted with sample polishing and preparation. K.V.G., S.E., A. D. W., G.H.A. and C.J. assisted with the measurements. J.V. and A.S.N. supervised the work. C.P.A., G.S., and J.V. wrote the manuscript with extensive input from the other authors. 

\vspace{-0.4cm}
\subsection*{Competing interests}
\vspace{-0.4cm}

The authors declare no competing interests.

\vspace{-0.4cm}
\subsection*{Corresponding author}
\vspace{-0.4cm}
Correspondence and requests for materials should be addressed to C. P. Anderson (cpand@illinois.edu) and J. Vu\v{c}kovi\'{c} (jela@stanford.edu).

\clearpage
\newpage
\mbox{~}

\SupplementaryMaterials
\setcounter{equation}{0}

\title{Supplemental information: Quantum critical electro-optic and piezo-electric nonlinearities}
\maketitle

\renewcommand{\thesection}{S\arabic{section}} 
\setcounter{section}{0}  
\begin{bibunit}

\vspace{-0.4cm}
\section{Miller's rules}
\label{sec:SIMiller}
\vspace{-0.4cm}

We begin with Miller's rules for second and third order optical nonlinearities \cite{Boydtextbook,byer1974nonlinear}:
\begin{equation}
    \chi^{(2)}(\omega_1,\omega_2,\omega_3)=\delta~\chi^{(1)}(\omega_1) \chi^{(1)}(\omega_2) \chi^{(1)}(\omega_3)
    \label{Millers2}
\end{equation}

\begin{multline}
    \begin{gathered}
    \chi^{(3)}(\omega_1,\omega_2,\omega_3,\omega_4)= \\
    \delta~ \chi^{(1)}(\omega_1) \chi^{(1)}(\omega_2) \chi^{(1)}(\omega_3) \chi^{(1)}(\omega_4)
    \label{Millers3}
    \end{gathered}
\end{multline}

In $ABO_3$ ferroelectrics \cite{10.1063/1.1657458} and anharmonic potentials\cite{Boydtextbook,byer1974nonlinear}, the proportionality constant, ``Miller's $\delta$" is found to be proportional to the electrical polarization of the crystal $P$. For electro-optics, where two fields are at optical frequencies $\omega_0$ (input and output) and the electric field is near DC ($\omega=0$) and by using the definition $n^2 = 1+\chi^{(1)}(\omega)$, then:

\begin{equation}
    \chi^{(2)}_{EO} = \delta~ (n^2-1)^2 (\epsilon_r-1) \approx n^4\epsilon_r 
    \label{Millers2}
\end{equation}

\begin{equation}
    \chi^{(3)}_{EO}= \delta~(n^2-1)^2 (\epsilon_r-1)^2 \approx n^4\epsilon_r^2
    \label{Millers3}
\end{equation}

Using the proportionality of $\delta$ to $P$:
%%%
\begin{equation}
    \chi^{(2)}_{EO} \propto n^4 P\epsilon_r
    \label{Millersp2}
\end{equation}

\begin{equation}
    \chi^{(3)}_{EO} \propto n^4 P\epsilon_r^2
    \label{Millersp3}
\end{equation}

\begin{figure}[htb!] 
\begin{center}
\includegraphics[width=1\columnwidth]{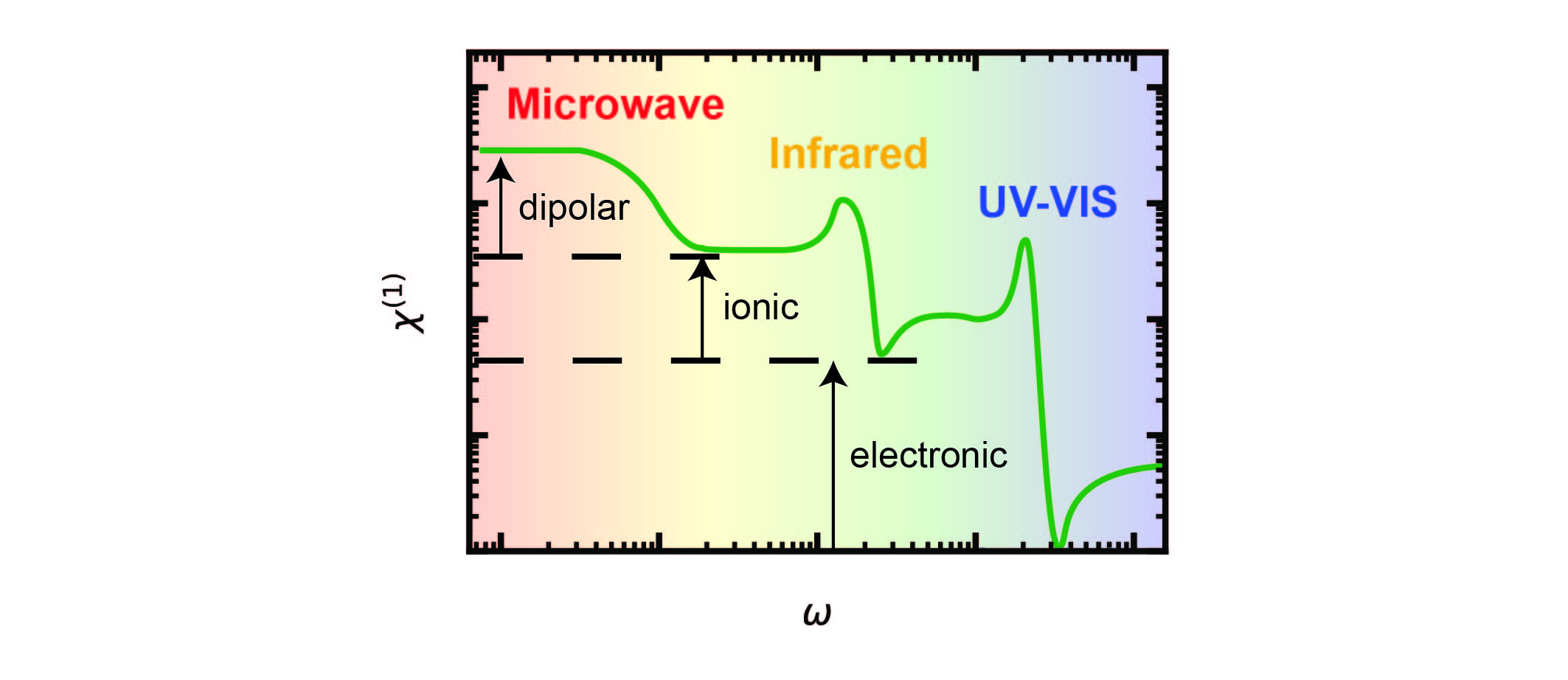}
\caption{
\textbf{Frequency dependent susceptibility $\vert$} The susceptibility $\chi^{(1)}$ is dependent on the electromagnetic frequency $\omega$, and is greatly enhanced in the microwave regime.
}
\label{fig:s8}
\end{center}
\vspace{-0.5cm}
\end{figure}

As many materials have a similar order of magnitude for the spontaneous electrical polarization $P$ in ferroelectrics (an ion mediating polarization can only move order one lattice constant), usually Miller's rules are framed purely in terms of susceptibilities. Importantly, because the optical susceptibilities are low and similar material-to-material, the low frequency $\chi^{(1)}$ is a major determining factor of nonlinearity (Fig. \ref{fig:s8}).

%%%%%%%%%%%%%%%%%%%%%%%%%%%%%%%%%%%%%%%%%%%%%%%%%%%%%%%
%%%%%%%%%%%%%%%%%%%%%%%%%%%%%%%%%%%%%%%%%%%%%%%%%%%%%%

\vspace{-0.4cm}
\section{Biased $\chi^{(2)}$}
\label{sec:SIbiasednon}
\vspace{-0.4cm}

An effective $\chi^{(2)}$ can be engineered by biasing a $\chi^{(3)}$ material with electric field. We first begin with the definitions of the nonlinear susceptibilities:
\begin{equation}
        P = \epsilon_0 \left( \chi^{(1)} E + \chi^{(2)} E^2 + \chi^{(3)} E^3 + \dots \right)
    \label{polarization expansion}
\end{equation}

Here, we can see that if $E=E_b+E_s$:

\begin{multline}
    \begin{gathered}
    P = \epsilon_0 \left( \chi^{(1)} (E_b+E_s)\right) + \\
    \epsilon_0 \left(\chi^{(2)} (E_b^2+2 E_b E_s + E_s)^2 + \dots \right)
    \label{polarization expansion}
    \end{gathered}
\end{multline}

Expanding:

\begin{multline}
    \begin{gathered}
    P_b+\epsilon_0(\chi^{(1)} E_s + \chi^{(2)} (2E_b E_s +E_s^2) \\
    +\chi^{(3)} (3E_b^2 E_s+3E_bE_s^2 +E_s^3)+\dots
    \end{gathered}
    \label{expansion2}
\end{multline}

Terms can be gathered according to the expansion with the signal field, $E_s$, such that:

\begin{equation}
    \chi^{(2)}_{eff} \approx \chi^{(2)}_0 + 3 E_b \chi^{(3)}_0
    \label{biased chi2}
\end{equation}

Where $\chi^{(2)}_0$ is the zero-bias $\chi^{(2)}$. Therefore, a DC bias can introduced a effective $\chi^{(2)}$, based on the strength of $\chi^{(3)}$ and the bias field, $E_b$. This has been utilized for cryogenic modulators in Si and SiGe , for example\cite{Chakraborty:20,Orcutt_2020}.

%%%%%%%%%%%%%%%%%%%%%%%%%%%%%%%%%%%%%%%%%%%%%%%%%%%%%%%
\begin{figure}[htb!] 
\begin{center}
\includegraphics[width=1\columnwidth]{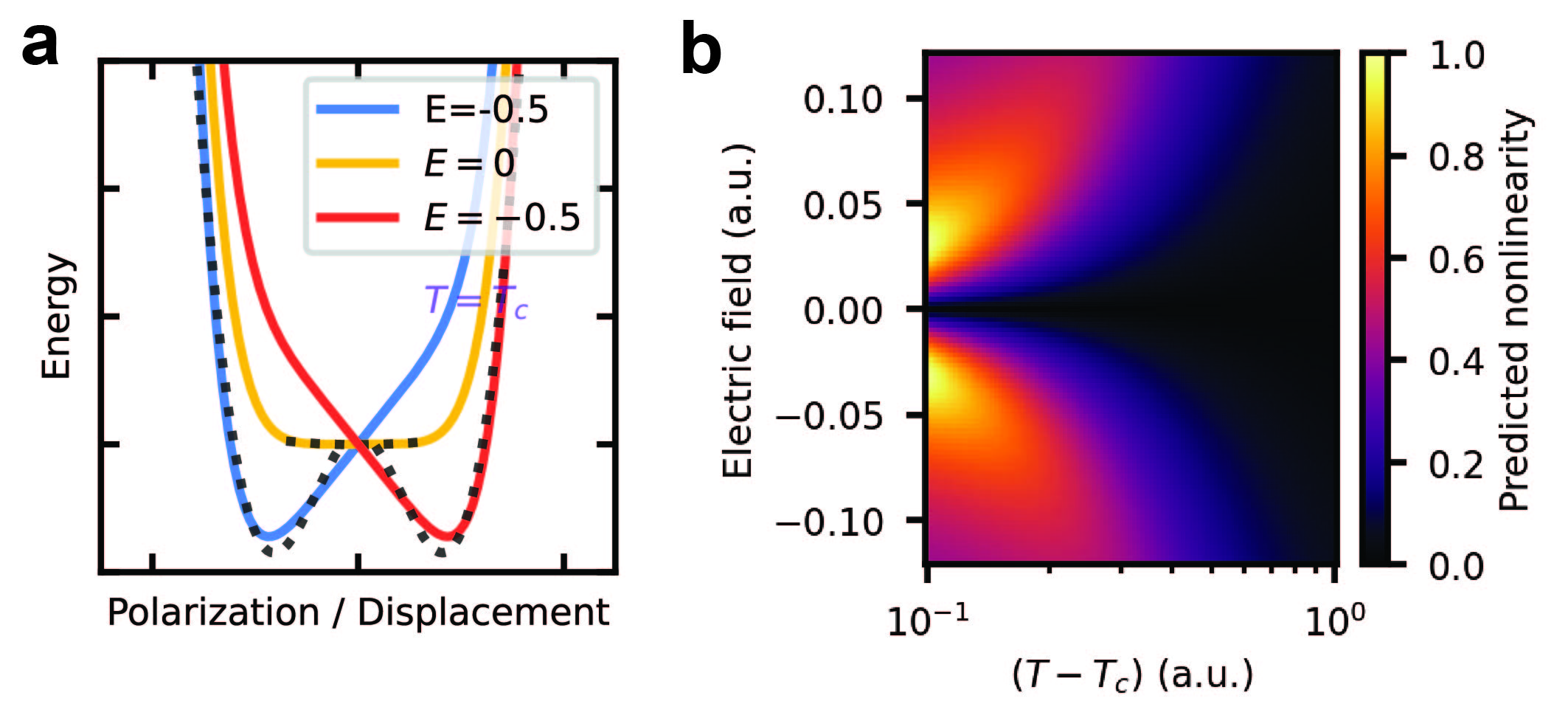}
\caption{
\textbf{Temperature and field dependent Landau-Ginzburg-Devonshire (LGD) model $\vert$} \textbf{a}, The Gibbs free energy landscape in the LGD model is affected by an external field $E$. This is shown at $T=T_{c}$. The new lowest energy states have nonzero $P$, and have induced anharmonicity (gray dotted line) which is a fit to a third order polynomial at the new equilibrium. \textbf{b}, Numerically calculated field and temperature dependent predicted nonlinearity from  ($P\epsilon_r$) with the LGD theory.
}
\label{fig:s9}
\end{center}
\end{figure}
%%%%%%%%%%%%%%%%%%%%%%%%%%%%%%%%%%%%%%%%%%%%%%%%%%%%%%

\vspace{-0.4cm}
\subsection*{Biased quadratic nonlinearities}
\vspace{-0.4cm}

Generally, linear electro-optic and piezo-electric effects in materials can be thought of as a quadratic nonlinearity biased by spontaneous, remnant, or induced electrical polarization\cite{10.1063/1.1657458,KNO,Newnham}. Beginning with the universal form of the associated quadratic nonlinearities:

\begin{equation}
    \Delta(\frac{1}{n^2})_{i}=g_{ij}P_{i}P_{j}
    \label{electrooptic}
\end{equation}

\begin{equation}
    x_{i}=q_{ij}P_{i}P_{j}
    \label{electrostriction}
\end{equation}

Where $\Delta(\frac{1}{n^2})$ is the change in the optical index matrix, and $x$ is the strain. For reference, if the change in index is small: 

\begin{equation}
    \Delta \left( \frac{1}{n^2} \right)_{i} \approx -\frac{2}{n_{i}^3} \Delta n_{i}
    \label{TaylorIndex}
\end{equation}

Specifically for $ABO_3$ perovskites, the polarization describes the motion of the $B$ ion, and the $B-O$ bond lengths directly change the bandgap of the crystal, modulating $n$. For linear dielectrics where $P=(\epsilon_{r}-1)\epsilon_0 E\approx\epsilon_{r}\epsilon_0 E$ and there is no built-in polarization, equations \ref{electrooptic} and \ref{electrostriction} become, in Voigt notation:

\begin{equation}
    \Delta(\frac{1}{n^2})_{i}=g_{ij}\epsilon_{r}^2\epsilon_0^2 E_{j}^2
    \label{quadratic EO}
\end{equation}

\begin{equation}
    x_{i}=q_{ij}\epsilon_{r}^2\epsilon_0^2 E_{j}^2
    \label{quadratic piezo}
\end{equation}

%%%%%%%%%%%%%%%%%%%%%%%%%%%%%%%%%%%%%%%%%%%%%%%%%%%%%%%
\begin{figure*}[htb!] 
\begin{center}
\includegraphics[width=2\columnwidth]{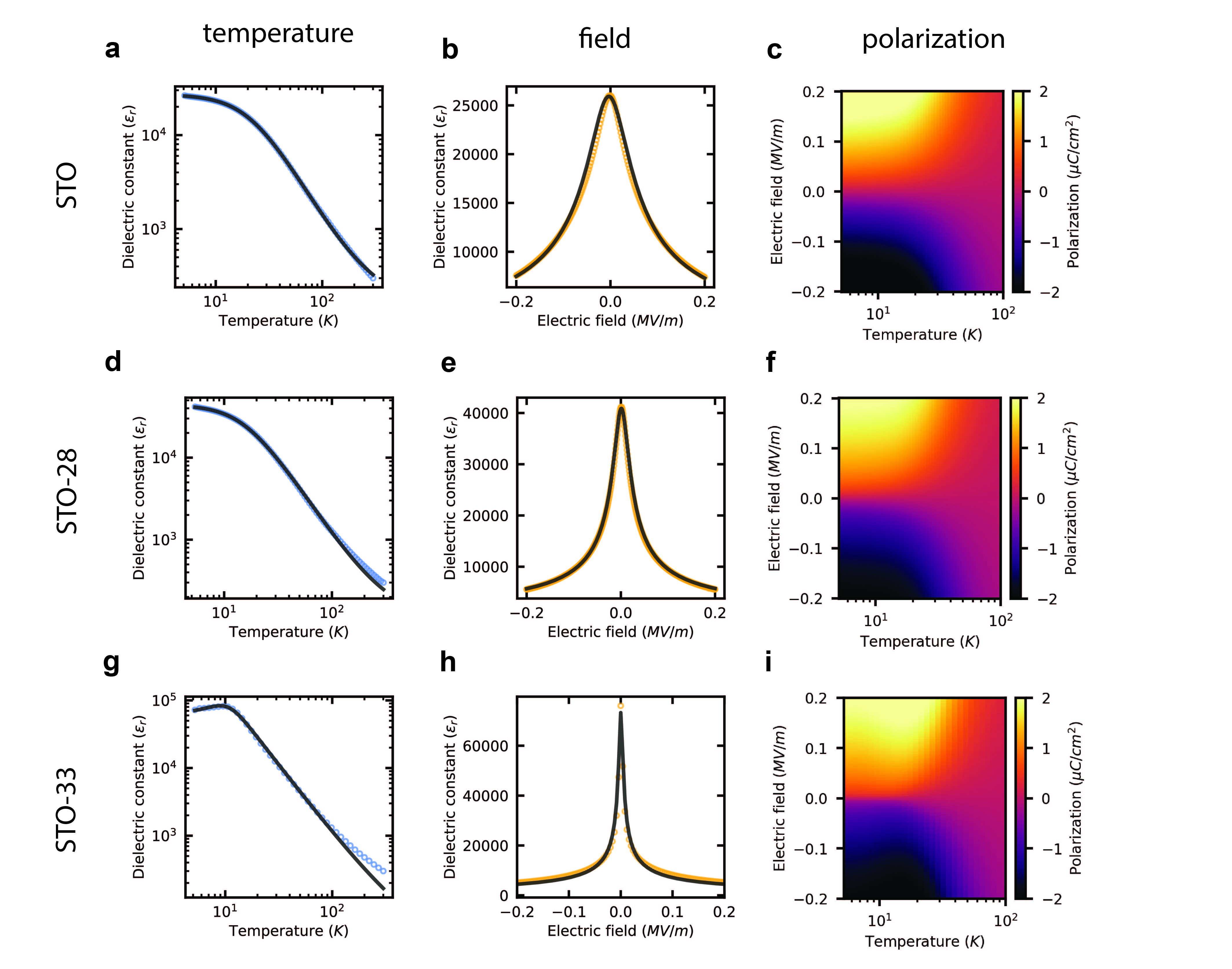}
\caption{
\textbf{Comparison of field, temperature, dependent dielectric and polarization data for natural and isotopically engineered STO  $\vert$} \textbf{a}, Temperature dependent dielectric data (blue circles), and fit (black line) at zero bias for STO. \textbf{b}, Electric field dependent dielectric data (yellow circles) and fit (black line) for STO at $T=5$ K \textbf{c}, Experimentally calculated electrical polarization as a function of bias field and temperature. \textbf{d},\textbf{e}.\textbf{f} and \textbf{g},\textbf{h}.\textbf{i} mirror the descriptions for \textbf{a},\textbf{b}.\textbf{c} for STO-28 and STO-33, respectively.
}
\label{fig:s1}
\end{center}
\vspace{-0.5cm}
\end{figure*}
%%%%%%%%%%%%%%%%%%%%%%%%%%%%%%%%%%%%%%%%%%%%%%%%%%%%%%

%%%%%%%%%%%%%%%%%%%%%%%%%%%%%%%%%%%%%%%%%%%%%%%%%%%%%%%
\begin{figure*}[htb!] 
\begin{center}
\includegraphics[width=2\columnwidth]{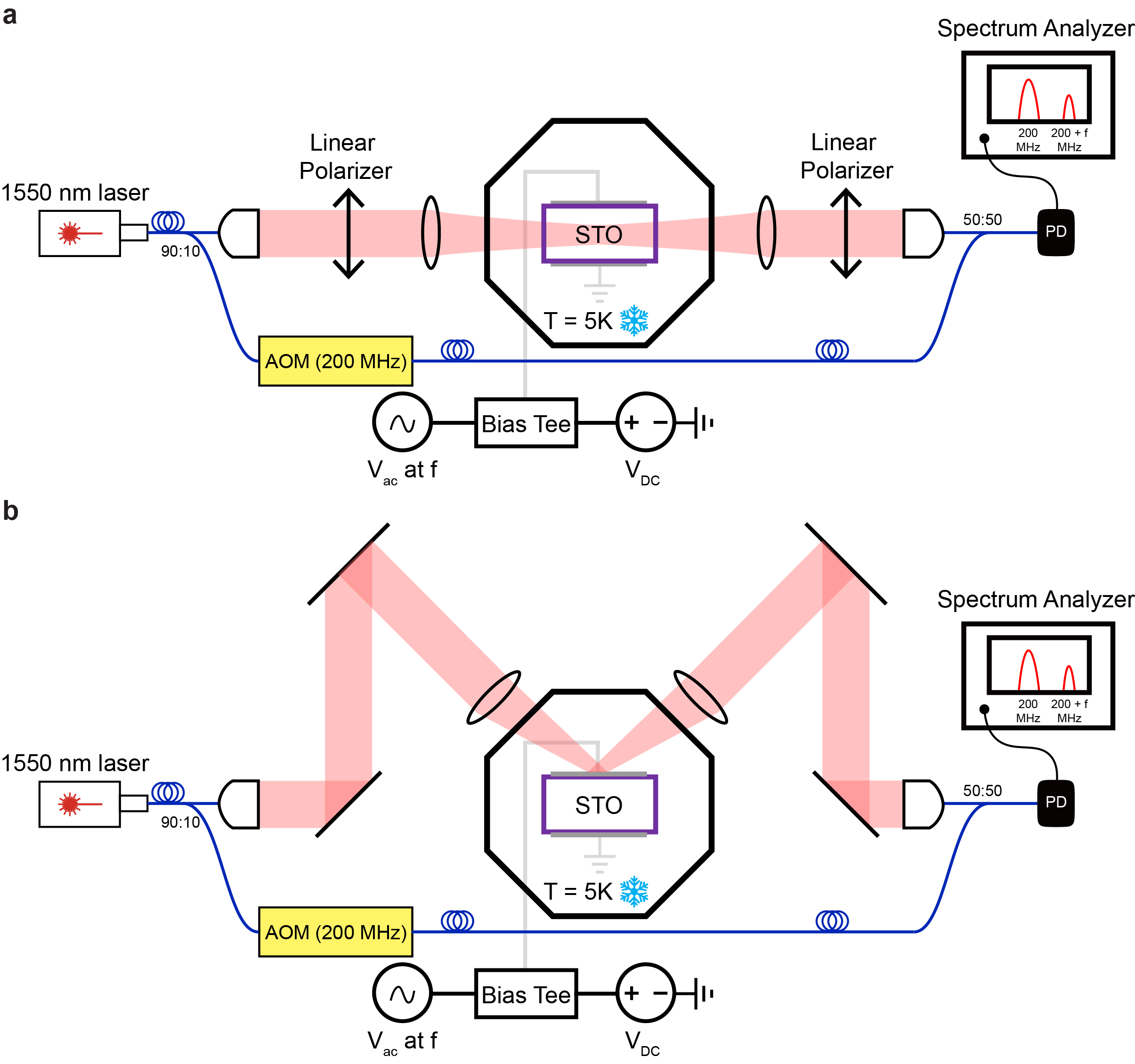}
\caption{
\textbf{Schematic of the experimental setup  $\vert$} \textbf{a}, Schematic of the experimental setup for electro-optic Pockels coefficient measurements. A 1550 nm laser is split into two arms of an MZI interferometer. In the first arm, the free-space beam passes through a polarizer and a lens which focuses the beam onto the STO sample. The beam propagates through the STO (driven by both AC and DC voltages), then is re-collimated by another lens and passes through a linear polarizer, which is set to the same direction as the first linear polarizer. The second arm contains a fiber-coupled AOM driven at 200 MHz. The two arms are recombined with a fiber splitter and the interference signal is measured by a fast photodiode (PD), which is connected to a spectrum analyzer. From the spectrum analyzer reading, the optical phase acquired by the sample beam is obtained, and the resulting Pockels coefficient for STO is calculated.  \textbf{b}, Schematic of the experimental setup for piezo-electric coefficient measurements. The setup is similar to \textbf{a}, except that the interferometer sample beam now reflects off the metalized surface of the STO, instead of transmitting through the material. This allows for the determination of the voltage-induced change in thickness of the STO and the resulting piezo-electric coefficient 
}
\label{fig:s4_exp}
\end{center}
\vspace{-0.5cm}
\end{figure*}
%%%%%%%%%

Here, we can see the well-known quadratic electro optic (Kerr) and quadratic piezo-electric (electrostrictive) effects with applied field. In this case, these third-order nonlinearities are proportional to $\epsilon_r^2$. Valid for both the paraelectric and ferroelectric state, the response of either an electrically biased system or a system with a spontaneous or remnant polarization can be framed as a biasing polarization ($P_b$) and a small signal polarization ($P_s$), such that $P=P_b +P_s$. Therefore, $P_{i}P_{j}=P_{b,i}P_{b,j}+P_{b,i}P_{s,j}+P_{b,j}P_{s,i}+P_{s,i}P_{s,j}$. With the signal small, and ignoring terms independent of the signal, $P_{i}P_{j}\approx P_{b,i}P_{s,j}+P_{b,j}P_{s,i}$. Therefore, if $i=j$, $P_{i}^2 \approx 2 P_{b,i}P_{s,i}$. This is the case in our experiment. We assume the linearity of small signals such that $P_{s}=(\epsilon_{r}-1)\epsilon_0 E_{s,j}$. From the definition of Pockels coefficient $r$ as $\Delta(\frac{1}{n^2})_{i}=r_{ij}E_{j}$ and the (inverse) piezo-electric coefficient $d$ as $x_{i}=d_{ij}E_{j}$, then:

\begin{equation}
    r_{ij}=2 g_{ij} (\epsilon_{r}-1)\epsilon_0 P_{b,j} \approx 2 g_{ij}\epsilon_{r}P_{b,j}
    \label{pockels}
\end{equation}

\begin{equation}
    d_{ij}=2 q_{ij}(\epsilon_{r}-1)\epsilon_0 P_{b,j}\approx 2 q_{ij} \epsilon_{r}P_{b,j}
    \label{piezo}
\end{equation}

These relations should hold for materials where ionic displacement mediates the nonlinearity (Fig. \ref{fig:s8}). On the other hand, purely electronic components of both nonlinearities exist (usually in non-ferroelectrics), but are in general weaker and not described here \cite{Hamze2020}.

Example LGD simulations are shown for a model ferroelectric in Fig. \ref{fig:s9}, showing that large changes in polarization are possible as a function of applied bias, mediating the nonlinearity in STO, for example. This is explored further in Fig. \ref{fig:s9}b, which shows the predicted nonlinearity ($P\epsilon_r$) as a function of temperature and bias above $T_c$. We see that the nonlinearity diverges as $T$ approaches $T_c$, and is maximized at a particular bias field. Thus, such electrically biased paraelectric materials near the phase transition remain a largely unexplored but promising\cite{Ren2002} route to tunable nonlinearity. 

\vspace{-0.4cm}
\section{Dielectric and polarization fitting}
\label{sec:SIpol}
\vspace{-0.4cm}

We show the temperature and voltage dependent line cuts of our fit for all three samples in Fig. \ref{fig:s1}. The calculated experimental polarization is shown, which is about $2~\mu C/cm^2$ at moderate bias for all three samples. This is understood as the the isotope substitution not changing the relevant length scale by which the $Ti^{4+}$ ion moves (polarization). The relevant features become sharper in electric field in STO-28 and STO-33. We note that our applied DC bias are more than two orders of magnitude below the dielectric breakdown of STO, which is roughly temperature independent at approximately 50 MV/m \cite{barrettbreakdown}. Therefore, more extreme bias-induced behavior of STO can be probed in integrated devices.

STO-33 is in close proximity to the QCP from previous works, and indeed shows a slight dielectric peaking (Fig. \ref{fig:s1}g) with temperature and hysteresis indicative of a small degree of ferroelectricity at $T=5$ K.

\vspace{-0.4cm}
\section{Details of the Experimental Setup}
\label{sec:SIMZI}
\vspace{-0.4cm}

Our Mach Zehnder interferometry (MZI) setup is schematically depicted in Fig. \ref{fig:s4_exp}. A 1550 nm laser ($\boldsymbol{E(t)}=\boldsymbol{E_0} cos(\omega_{0}t)$) is split into two arms. In the first arm, the sample beam travels a distance $d$ through the STO, which is in a parallel plate capacitor configuration with top and bottom electrical contact pads applying an electric field perpendicular to the beam propagation direction and parallel (perpendicular) to the beam polarization for $r_{33}$ ($r_{13}$) measurements. Using a bias tee, we apply both a DC bias voltage ($E_{bias}=V_{DC}/L$, where $L$ is the sample thickness) and an AC voltage with frequency $\Omega$: $V(t)=\frac{V_{AC}}{2}\sin(\Omega t)$. Because of the refractive index modulation from the electro optic effect, the sample beam acquires a time-varying phase $\phi(t)$ at frequency $\Omega$, and the transmitted field through the sample arm becomes:

\begin{equation}
    \boldsymbol{E_1(t)}=\boldsymbol{E_1} \cos{(\omega_{0}t + \phi_{0} \sin{(\Omega t))}}
    \label{sampleArm}
\end{equation}

The second arm of the MZI includes an acousto-optic modulator (AOM) driven at 200 MHz, which imparts a frequency shift of $\Delta \omega = 2\pi \times 200$ MHz on the propagating beam, resulting in a field:

\begin{equation}
    \boldsymbol{E_2(t)}=\boldsymbol{E_2} \cos{((\omega_{0}+\Delta\omega)t)}
    \label{AOMarm}
\end{equation}

The two arms of the interferometer are then recombined, and the total intensity signal is measured with a photodiode:

\begin{align}
    I(t)\propto |E_{total}|^2 = |E_1(t)+E_2(t)|^2 \\
    I(t)\propto E_1(t)^2+E_2(t)^2+2E_1(t)E_2(t)
    \label{IntensityInt}
\end{align}    

In particular, the interference term:

\begin{equation}
    2E_1(t)E_2(t) \propto \cos{(\Delta\omega t -\phi_{0}\sin{(\Omega t)})}
    \label{Int term}
\end{equation}

results in a detected signal with components at the beat frequency of 200 MHz and sidebands at $200$ MHz$\pm \Omega$. The photodiode outputs a current onto a 50$\Omega$ load that is proportional to the intensity of the optical signal. We connect the photodiode output to the RF + DC port of a bias tee, which is 50$\Omega$ terminated on the DC port while the RF port is connected to an electronic spectrum analyzer (ESA, with 50$\Omega$ input impedance). The ESA records the powers $P_0$ and $P_1$ of the 200 MHz and at $200$ MHz$+\Omega$ signals. This directly provides information about the phase $\phi_0$ from the STO, as can be seen by taking the Jacobi-Anger expansion of the interference term in Eq.~\ref{Int term} above: 

\begin{equation}
    \cos[\Delta \omega t - \phi_0 \sin(\Omega t)] = \sum_{n=-\infty}^\infty J_n(\phi_0) \cos[(\Delta \omega - n\Omega) t]
    \label{Jacobi}
\end{equation}

where $J_n(x)$ is the Bessel function of the first kind. Here, we see that the $n=0$ term is responsible for the signal at the carrier (200 MHz) frequency, and the $n=\pm1$ terms correspond to the signal at the first sidebands. Specifically, the intensities of the carrier ($A_0$) and ($A_{\pm}$) signals are proportional to $J_0(\phi_0)$ and $J_{\pm1}(\phi_0)$, respectively:

\begin{equation}
    \frac{A_{\pm1}}{A_0} = \frac{J_{\pm1}(\phi_0)}{J_0(\phi_0)}
    \label{Amplituderatio}
\end{equation}

In our experiments the phase modulation is small $\phi_0<<1$ so we can approximate the Bessel functions:

\begin{align}
    J_0(\phi_0) \approx 1 \\
    J_1(\phi_0) \approx \frac{\phi_0}{2}
    \label{smallangleBessel}
\end{align} 

Rearranging Eq.~\ref{Amplituderatio} we obtain:

\begin{equation}
    \phi_0 \approx 2 \frac{A_{\pm1}}{A_0}.
    \label{phase_amplitude}
\end{equation}

The spectrum analyzer connected to the photodiode measures the carrier and sideband powers in dbm ($P_0$ and $P_{\pm1}$, respectively). Because the measured power scales as the square of the current outputted by the photodiode, the relationship between the measured power and output photodiode current (proportional to the interference intensity) is given by:

\begin{align}
    A_{0}=\sqrt{10^{P_{0}/10}} \\
    A_{\pm1}=\sqrt{10^{P_{\pm1}/10}}
    \label{PowertoAmplitude}
\end{align} 

Combining Equations~\ref{phase_amplitude} and~\ref{PowertoAmplitude}, we obtain:

\begin{equation}
    \phi_0 = 2\sqrt{10^{(P_{\pm1}-P_{0})/10}}.
    \label{phase_power}
\end{equation}

%%%%%%%%%%%%%%%%%%%%%%%%%%%%%%%%%%%%%%%%%%%%%%%%

%%%%%%%%%%%%%%%%%%%%%%%%%%%%%%%%%%%%%%%%%%%%%%%%%%%%%%

\vspace{-0.4cm}
\subsection*{Extraction of the Pockels coefficient from the MZI experiment}
\vspace{-0.4cm}

As explained in the main text and above (Equations~\ref{electrooptic} and \ref{TaylorIndex}), the electro-optic effect is given by: 

\begin{equation}
    \Delta n_{i}=-\frac{1}{2} {{n}_{i}^3} r^{eff}_{ij} E^{AC}_{j}
    \label{Pockels_eff}
\end{equation}

Where $r_{ij}^{eff}$ is defined in Eq.~\ref{pockels}. In our experiments we focus primarily on the $r_{33}\equiv r^{eff}_{33}$ tensor elements, which we pick out experimentally by polarizing the optical signal in the same direction as the applied DC and AC electric fields.  Rearranging Eq.~\ref{Pockels_eff}, we obtain:

\begin{equation}
    r_{33}=-\frac{2\Delta n}{{n}^3 E_{AC}}
    \label{r_1}
\end{equation}

We note that the same treatment applies for the $r_{13}\equiv r^{eff}_{13}$ term, with the only difference being that the optical signal is polarized in an orthogonal direction in the MZI compared to the $r_{33}$ measurements.

In our MZI measurement, we measure optical phase $\phi_0$ through the STO, which is related to $\Delta n$ by:

\begin{equation}
    \phi_0=2\pi \frac{\Delta n d}{\lambda}
    \label{phase}
\end{equation}

%%%%%%%%%%%%%%%%%%%%%%%%%%%%%%%%%%%%%%%%%%%%%%%%%%%%%%%
\begin{figure*}[htb!] 
\begin{center}
\includegraphics[width=2\columnwidth]{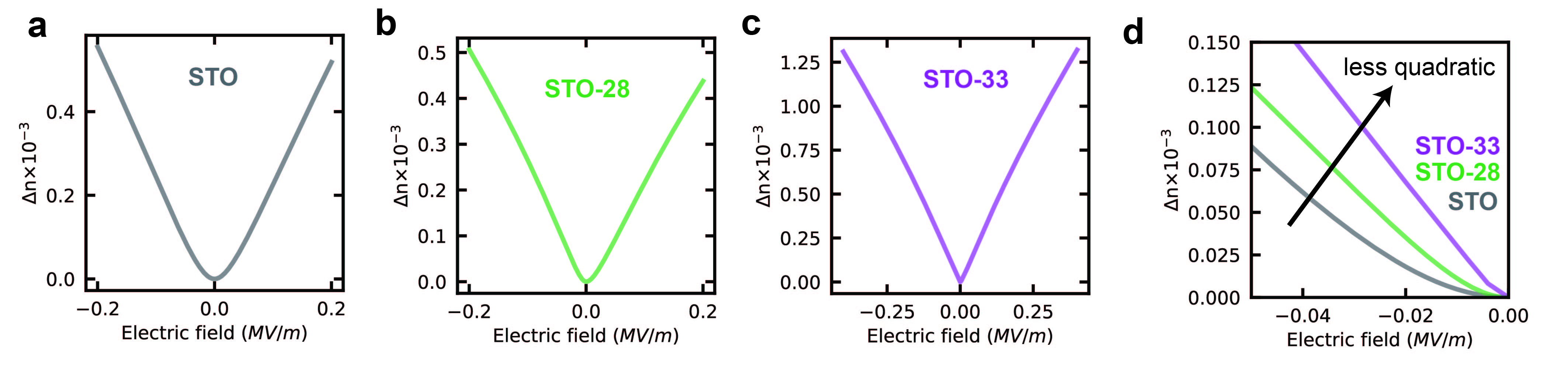}
\caption{
\textbf{Changes in the refractive index with electric field in STO  $\vert$} \textbf{a}, Measured index change for STO. \textbf{b}, Measured index change for STO-28. \textbf{c}, Measured index change for STO-33. \textbf{d}, Zoom-in to the low field regime for all three samples, as the isotope fraction is tuned, the slope becomes higher and the shape becomes less quadratic and more linear.
}
\label{fig:s14}
\end{center}
\vspace{-0.5cm}
\end{figure*}
%%%%%%%%%

Where $d$ is the distance the optical signal travels through the STO between the electrodes, and $\lambda$ is the wavelength (1550 nm in our experiments).
Thus:
\begin{equation}
    r_{33}=-\frac{\lambda \phi_0}{\pi d n^3 E_{AC}}
    \label{r_2}
\end{equation}

Ignoring the minus sign, substituting for $E_{AC}=V_{AC}/(2L)$ (since $V_{AC}$ is the peak-to-peak voltage) and Eq.~\ref{phase_power}:

\begin{equation}
    r_{33}=\frac{2 \lambda L \sqrt{10^{(P_{\pm1}-P_{0})/10}}}{\pi n^3 d \frac{V_{AC}}{2}}
    \label{r_final}
\end{equation}

\vspace{-0.4cm}
\subsection*{Extracted values of $r_{13}$}
\label{sec:SIr13}
\vspace{-0.4cm}

The measured ratio between $r_{33}$ and $r_{13}$ is $\approx$3.33 ($r_{13}$=170(20) pm/V), consistent with literature reports of a ratio between 2.8 and 4 at room temperature \cite{Geusic,Fujiicoeff}. 

For STO-33, the ratio between $r_{33}$ and $r_{13}$ also grows to about 4.5 (versus about 3 for natural STO). This may be related to the increased anisotropy and birefringence near the quantum phase transition.

\vspace{-0.4cm}
\subsection*{Extraction of the piezo-electric coefficient from the MZI Experiment}
\vspace{-0.4cm}

We measure the piezo-electric coefficient $d_{33}$ with the MZI setup similar to the one used for Pockels coefficient, except that the sample beam is reflected off the top (reflective) surface of the STO, at an angle of 45$^\circ$. This enables a measurement of the change in thickness of the STO chip for a given applied (AC) voltage. 

The piezo-electric coefficient is given by:

\begin{equation}
    d_{ij}=\frac{x_i}{E_j}
    \label{d_def}
\end{equation}

Where $x_i$ is the mechanical strain, defined as:

\begin{equation}
    x_i=\frac{\Delta L}{L}
    \label{S_def}
\end{equation}

In our measurements, the electric field is applied to the STO in a parallel plate capacitor configuration (as in the Pockels coefficient measurements), so $L$ is the sample thickness, and $\Delta L$ is the change in thickness arising from the applied electric field. The change in sample thickness $\Delta L$ can be related to a change in optical phase $\phi_0$ by geometric arguments. Given the incidence angle of $45^\circ$, the phase acquired by sample beam is:

\begin{equation}
    \phi_0=2\pi \frac{n_{0} 2\Delta L}{\sin(45^\circ)\lambda} = \frac{4\sqrt{2} \pi \Delta L}{\lambda}
    \label{phi_thickness}
\end{equation}

Combining Equations~\ref{d_def},~\ref{S_def} and~\ref{phi_thickness} we obtain:

\begin{equation}
    d_{33}=\frac{\lambda \phi_0}{4\sqrt{2}\pi L E_{AC}}
    \label{d33_phi}
\end{equation}

As in the previous section, $E_{AC}=V_{AC}/2L$, since $V_{AC}$ is defined as the peak-to-peak voltage. Combining Eq.~\ref{phase_power} and Eq.~\ref{d33_phi}, we obtain:

\begin{equation}
    d_{33}=\frac{\lambda \sqrt{10^{(P_{\pm1}-P_{0})/10}}}{2\sqrt{2}\pi \frac{V_{AC}}{2}}.
    \label{d33_final}
\end{equation}

Finally, we note that because we take the ratio between the sideband and 200 MHz carrier signal, our measurements of both the Pockels coefficient and piezo-electric coefficient are independent of the responsivity of the detector, losses, power imbalance of the interferometer arms, and other sources of error. 

\vspace{-0.4cm}
\section{Frequency dependence}
\label{sec:SIfreq}
\vspace{-0.4cm}

%%%%%%%%%%%%%%%%%%%%%%%%%%%%%%%%%%%%%%%%%%%%%%%%%%%%%%%
\begin{figure}[H] 
\begin{center}
\includegraphics[width=1\columnwidth]{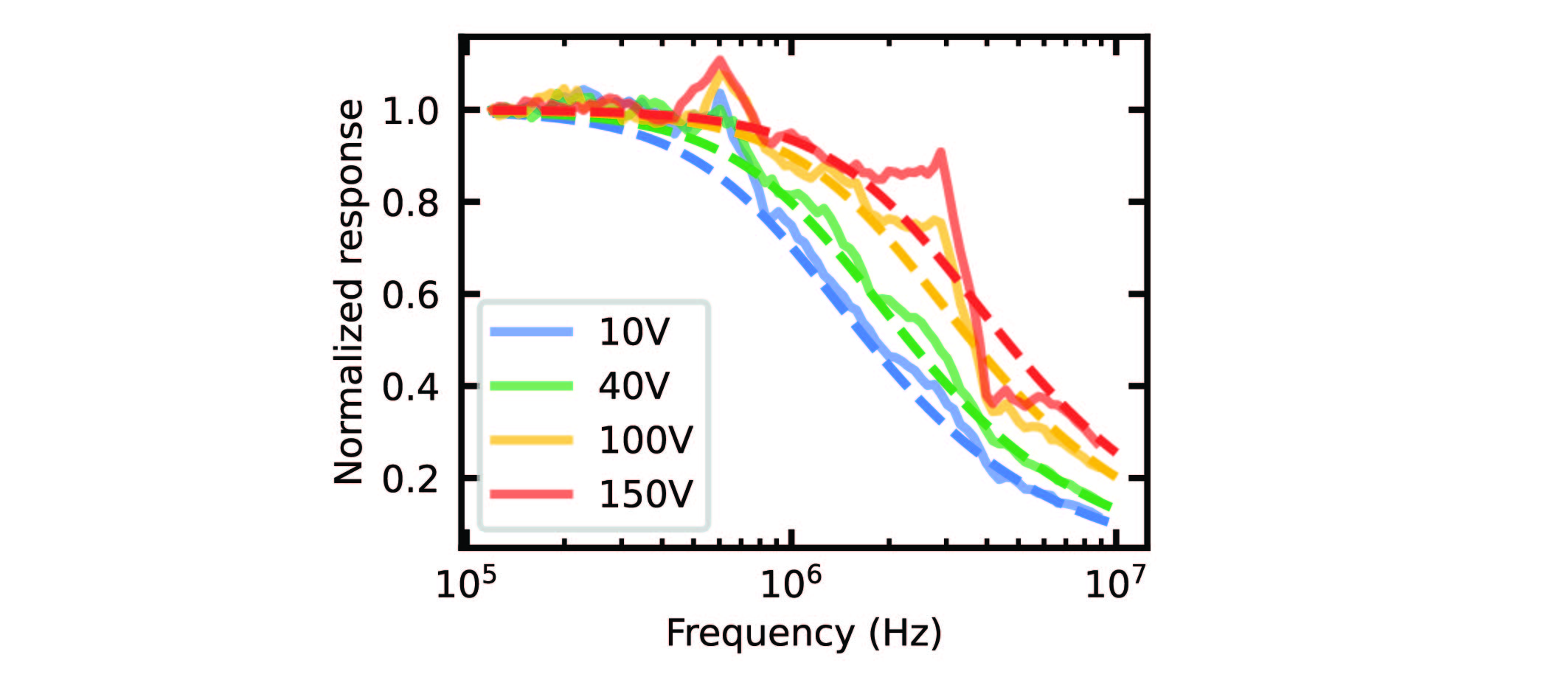}
\caption{ \textbf{Frequency dependent electro-optic response as a function of electrical bias for STO $\vert$} Data is normalized to the low frequency value for a variety of electrical voltages (shaded line). The dotted lines correspond to a prediction using the observed voltage dependent capacitance. Data is taken at $T=5$ K.
}
\label{fig:s3}
\end{center}
\end{figure}
%%%%%%%%%%%%%%%%%%%%%%%%%%%%%%%%%%%%%%%%%%%%%%%%%%%%%%

The voltage dropped on our sample, a capacitor, $V_C$ with a given applied $V$ in series with a resistance $R$ can be modeled as:

\begin{equation}
    \frac{V_C}{V}= \left| \frac{-j Z_c}{-j Z_c + R} \right|
\end{equation}

Where $Z_c = \frac{-j}{2 \pi f C}$. We measure the frequency dependence of the Pockels modulation at multiple different bias voltages for natural STO. At the different voltages, we use the measured capacitance and assume $R=50~\ohm$ to model the frequency dependence. Fig. \ref{fig:s3} shows a good agreement with our experimental data, where the 3 dB cutoff is pushed to higher frequencies at large bias fields. For our measurements, this is around 1-4 MHz. Therefore, the observed strength of modulation should not be limited significantly at 100 kHz and is entirely explained by electrical considerations, not a reduction in the Pockels effect at higher frequencies. The observed structure in our signal (Fig. \ref{fig:s3}) is not noise, but instead high frequency electrical and mechanical resonances in our sample that are also avoided at low frequency. We also observed a weaker frequency dependence of the electro-optic coefficient in LN, consistent with its orders of magnitude lower dielectric constant and higher RC cutoff.

\vspace{-0.4cm}
\section{Theoretical predictions of the electro-optic bandwidth}
\label{sec:SIEOBand}
\vspace{-0.4cm}

%%%%%%%%%%%%%%%%%%%%%%%%%%%%%%%%%%%%%%%%%%%%%%%%%%%%%%%
\begin{figure}[H] 
\begin{center}
\includegraphics[width=1\columnwidth]{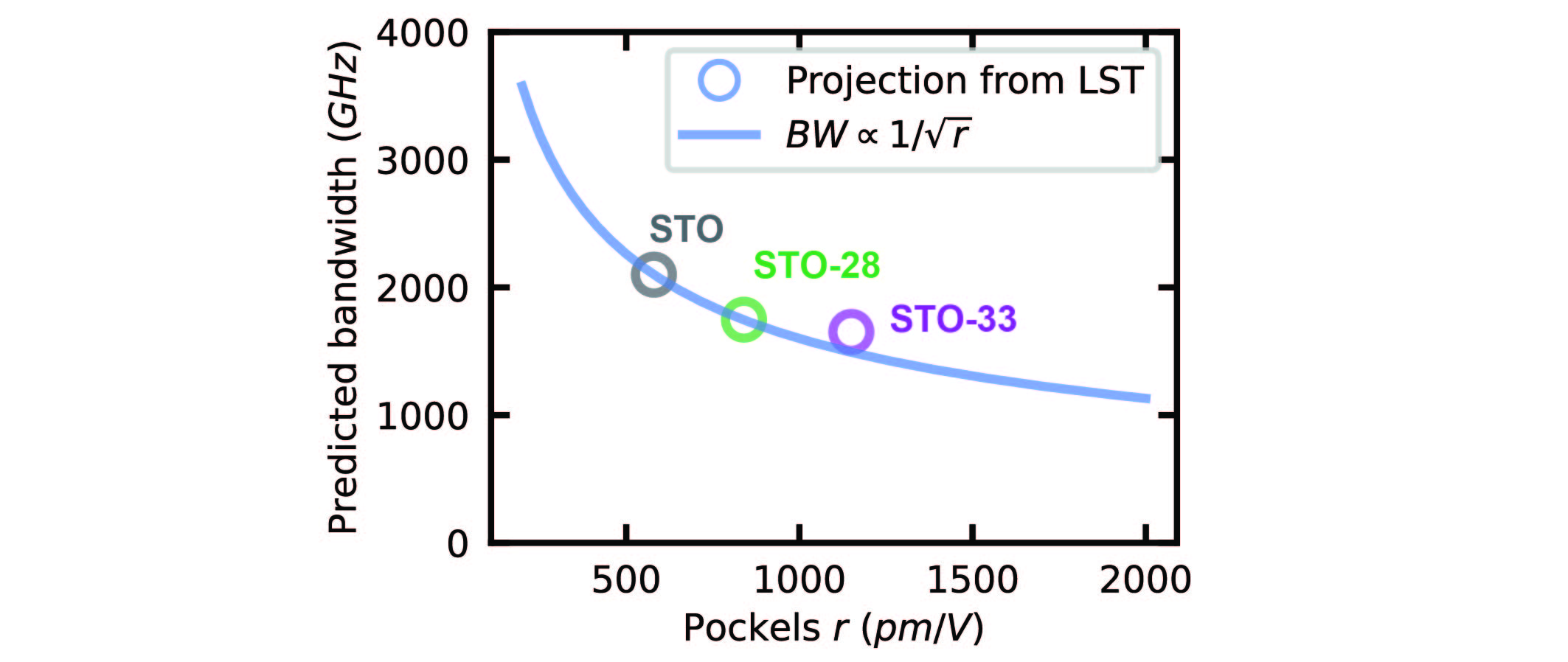}
\caption{ \textbf{Predicted electro-optic bandwidth $\vert$} The blue dots are found from the measured $r$, and the dielectric constant at that bias field $\epsilon_r(V_{max})$ using the LST relation and the known phonon mode frequencies. The solid line is a predicted scaling if $BW\propto 1/\sqrt{r}$. All values are at $T=5$ K.
}
\label{fig:s15}
\end{center}
\end{figure}
%%%%%%%%%%%%%%%%%%%%%%%%%%%%%%%%%%%%%%%%%%%%%%%%%%%%%%

The frequency dependence of the dielectric response (Fig. \ref{fig:s1}) sets the frequency dependence of the electro-optic response, as shown by Miller's rules and the results in this manuscript.

The relevant frequency scale where the enhanced dielectric properties of STO, and other ``soft" mode ferroelectrics is set by the relevant phonon frequency\cite{Neville}. There, the Lyddane-Sachs-Teller (LST) relation holds (including for STO), which states: 

\begin{equation}
 \frac{\epsilon(0)}{\epsilon(\infty)} = \frac{\omega_L^2}{\omega_T^2},
\end{equation}

where $\omega_T$ is the frequency of a long-wavelength transverse optic phonon and $\omega_L$ the frequency of the corresponding long-wavelength longitudinal optical phonon which mediates electrical polarization. $\epsilon(0)$ and $\epsilon(\infty)$ are the permittivities at frequencies below and above $\omega_T$\cite{Neville}.

The result of this relation is that the maximum frequency at which the electro-optic response is large (or the bandwidth ``BW") is $\approx \omega_T$ and that $\omega_T^2\propto\epsilon_r$. If, as we predict from Miller's rules and observe (Fig. \ref{fig:fig4}b) that $r\propto\epsilon_r$, then:

\begin{equation}
    BW \propto 1/\sqrt{r}
\end{equation}

The phonon frequencies are well known at low temperature for STO, at about 1.4 THz in the (zero bias) quantum paraelectric state \cite{phononPhysRevB.110.054302,STOphonon}. In fact, the reduction in dielectric constant with field exactly corresponds to a hardening of the phonons in this previous work.

We use this previous literature and insight to predict an intrinsic trade-off between the nonlinearity and operating bandwidth, shown in Fig. \ref{fig:s15} For all our samples, the theoretical BW limit is above 1 THz, using the observed ``hardened" dielectric value that shows maximum nonlinearity. Therefore, STO and quantum critical STO may find wide use in high frequency applications up to many hundred GHz. These trends should also hold for piezo-electricity. These calculations make no evaluation of RF losses, just fundamental material response time.

\vspace{-0.4cm}
\section{Change in index of refraction with field}
\label{sec:SIquadindex}
\vspace{-0.4cm}

Due to the nonlinear change of the electrical polarization with field, the electro-optic response transitions from quadratic to linear as a function of field. We plot the total change in index as a function of field for our three samples in Fig. \ref{fig:s14}. The samples tuned to criticality have a higher slope, but display less quadratic electro-optic behavior and increased linearity. \textit{STO is not a purely quadratic electro-optic material.}

\vspace{-0.4cm}
\section{Optical loss}
\label{sec:SILoss}
\vspace{-0.4cm}

The optical loss of SrTiO$_3$ is measured through the transmission of light through the bulk crystal with a power meter. The sample is moved within the cryostat with a cryogenic nanopositioner to either have the light pass through it, or bypass to the side. Therefore, a precise measurement of the loss purely due to the sample is obtained. The observed loss arises from the Fresnel reflections at the two polished interfaces, scattering due to roughness or metal, and intrinsic loss through the 5 mm crystal. Using the known index of refraction of STO, we correct for Fresnel reflections, resulting in an estimate of intrinsic loss at room temperature between 0-0.2 dB/cm. This is limited primarily by the power stability of the laser and should be considered an \textit{upper limit}.

%%%%%%%%%%%%%%%%%%%%%%%%%%%%%%%%%%%%%%%%%%%%%%%%%%%%%%%
\begin{figure}[H] 
\begin{center}
\includegraphics[width=1\columnwidth]{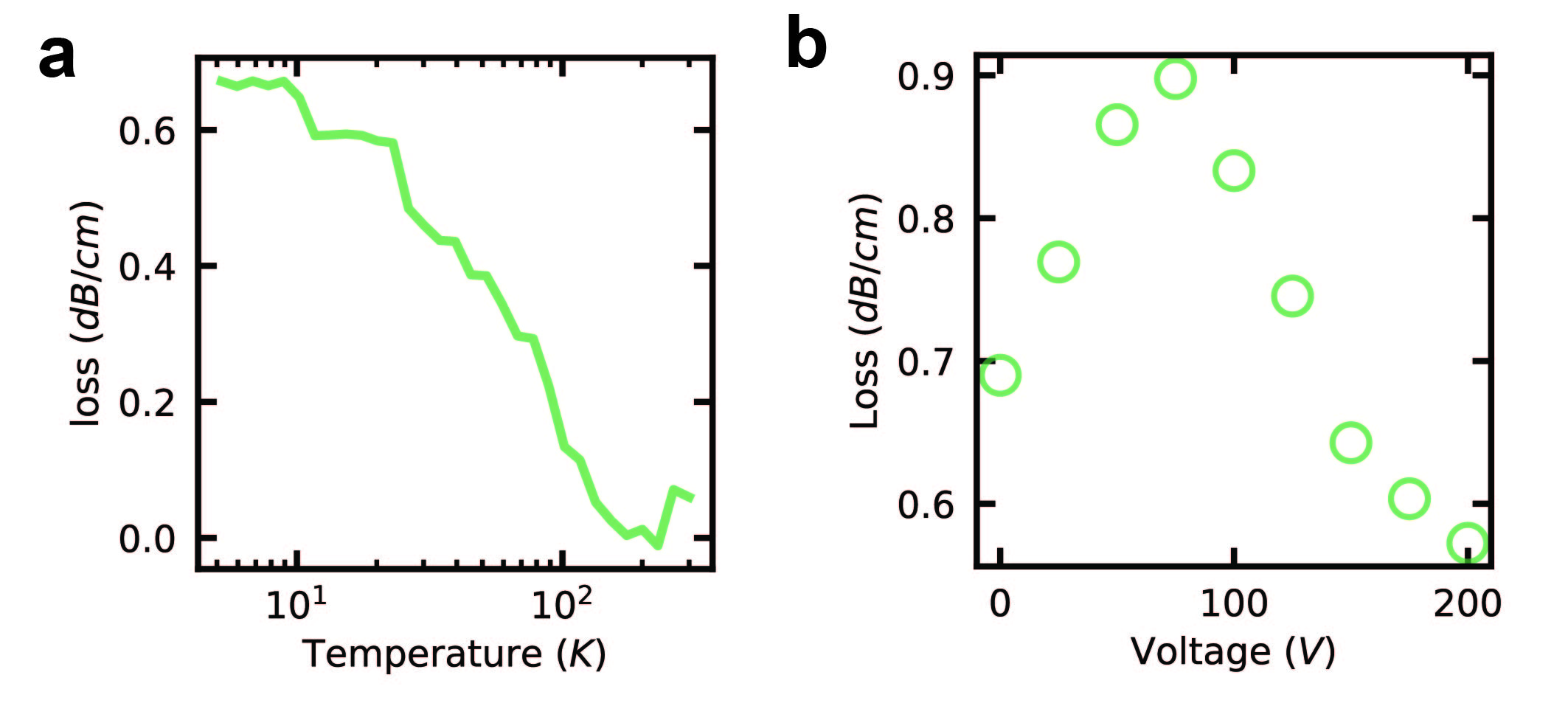}
\caption{\textbf{Optical loss in STO $\vert$} \textbf{a}, optical loss measured as a function of temperature. \textbf{b}, loss as a function of applied DC bias voltage at $T=5K$.
}
\label{fig:s4}
\end{center}
\end{figure}
%%%%%%%%%%%%%%%%%%%%%%%%%%%%%%%%%%%%%%%%%%%%%%%%%%%%%%

We plot the temperature dependence of the optical loss in Fig. \ref{fig:s4}a. As STO cools, the loss increases but saturates at about 0.7 dB/cm, mirroring the quantum paraelectric dependence of the dielectric constant. This is sufficient for most integrated photonics applications.

The loss also depends on the applied bias, shown in Fig. \ref{fig:s4}b. At zero bias, the loss is about 0.7 dB/cm, which then peaks to around 0.9 dB/cm at around 75 V (also where the nonlinearity is maximized), but then rapidly decays to lower than 0.55 dB/cm at 200 V bias . We hypothesize that this behavior is due to the formation of so-called polar nano-regions (PNRs)\cite{pnas}, that have random orientations that scatter light due to differences in the index of refraction. As the bias grows, however, the applied field strongly aligns the crystal, reducing inhomogeneity. There is therefore an interesting trade space between nonlinearity, loss, and bias, that required further investigation, especially in integrated optical devices with larger bias fields.

\vspace{-0.4cm}
\section{High resolution scans of the Pockels nonlinearity of STO-33}
\label{sec:SISTO33}
\vspace{-0.4cm}

\begin{figure}[htb!] 
\begin{center}
\includegraphics[width=1\columnwidth]{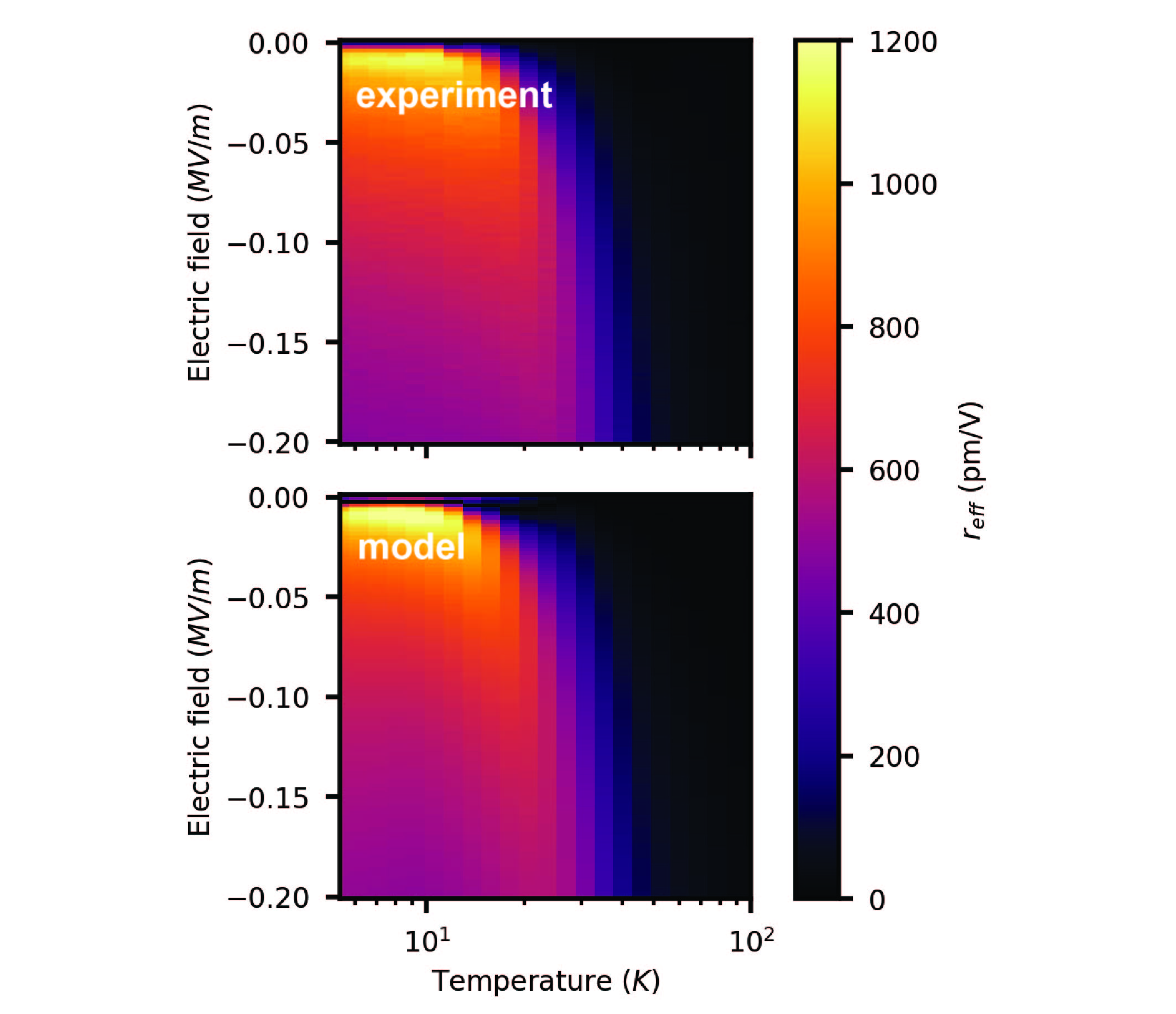}
\caption{
\textbf{Fine, single sided scan of the Pockels nonlinearity in STO-33 $\vert$} Observed Pockels effect and model using a single sided and more fine scan. This avoids hysteresis from the near-ferroelectric nature of the sample. This dataset is used to report the values in the main text.
}
\label{fig:s6}
\end{center}
\end{figure}

\begin{figure}[htb!] 
\begin{center}
\includegraphics[width=1\columnwidth]{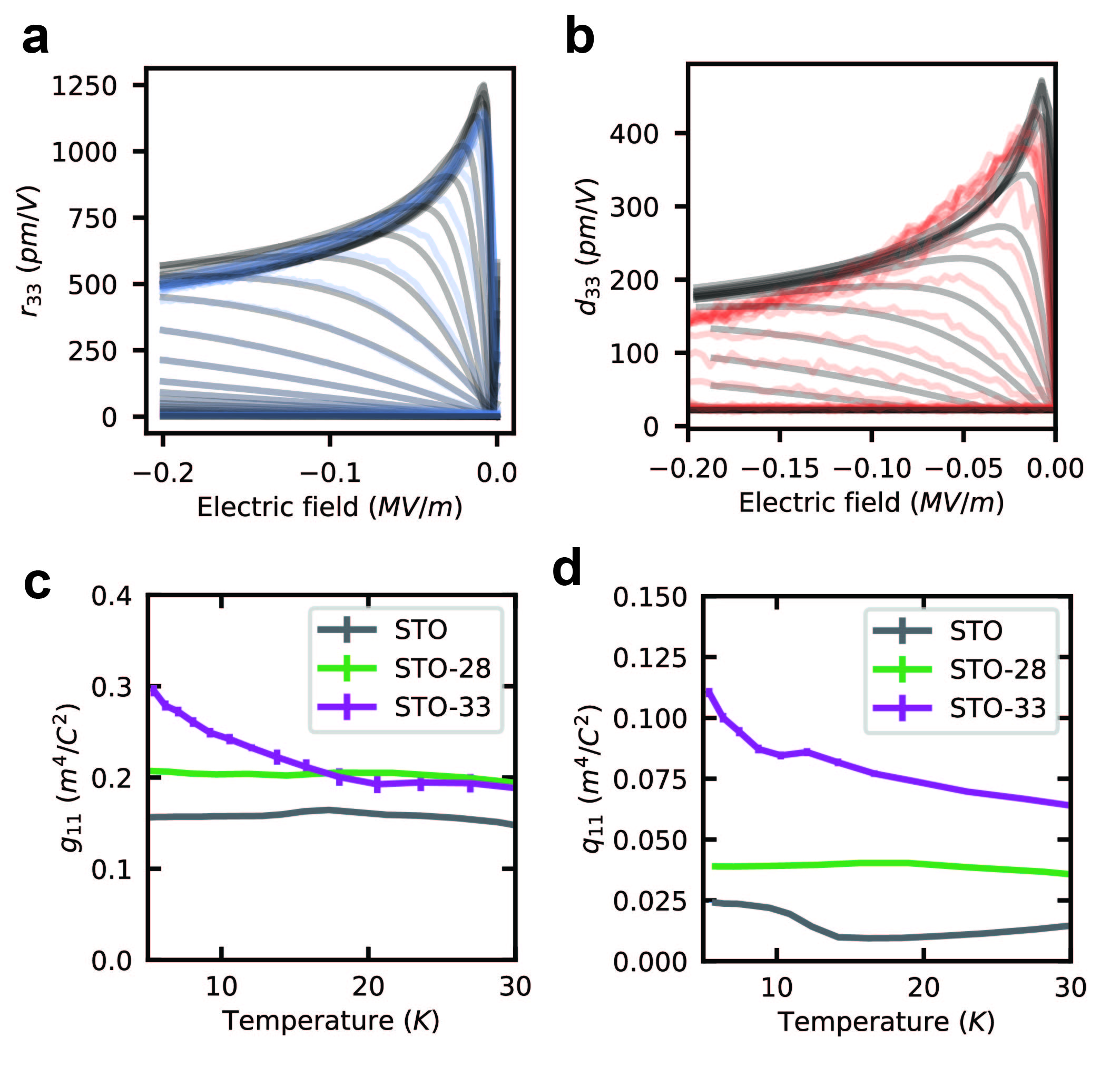}
\caption{
\textbf{Temperature dependent fitting of $r_{33}$ and $d_{33}$ for STO-33 and extracted $g_{11}$, $q_{11}$ for all samples $\vert$} \textbf{a}, Line by line fits (black) of the single sided electro-optic data (blue) and \textbf{b}, piezo-electric data (red) as a function of temperature. \textbf{c},\textbf{d}, Show the temperature dependence of the fitted $g_{11}$ and $q_{11}$ of STO, STO-28, STO-33.
}
\label{fig:s7}
\end{center}
\end{figure}

\begin{figure*}[htb!] 
\includegraphics[width=2\columnwidth]{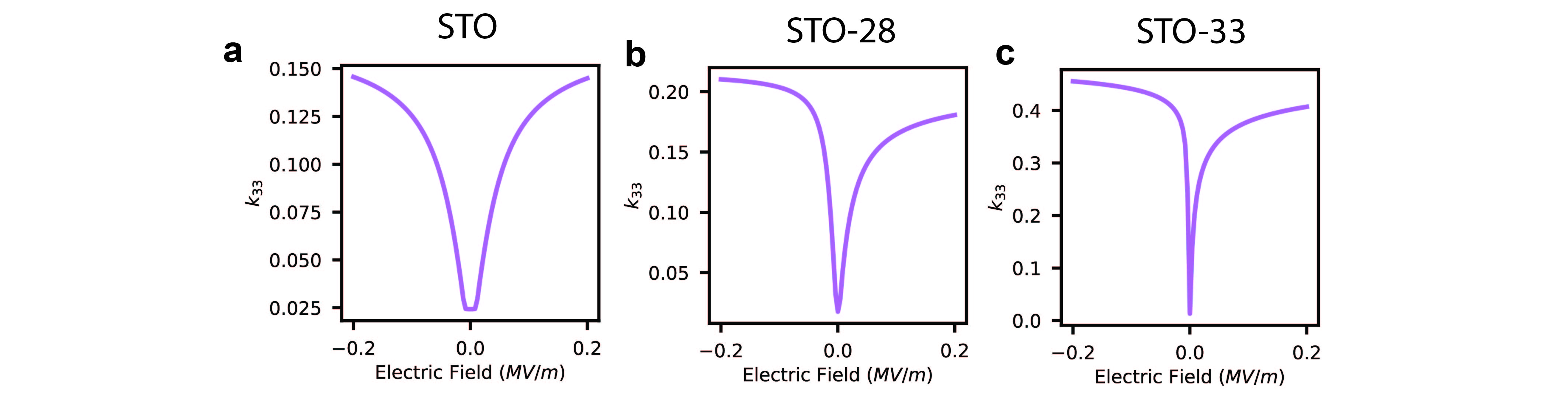}
\caption{\textbf{Electromechanical coupling across the three samples $\vert$} \textbf{a},\textbf{b},\textbf{c}, Show a electric field dependent line cut of $k_{33}$ at $T=5K$ for STO, STO-28, STO-33, respectively. The ``bottoming out" of (\textbf{a}) is the noise floor of our measurement.
}
\label{fig:s2}
\end{figure*}

As there are some sweep speed dependent and hysteric effects in STO-33 we also report a slow, high resolution scan without biasing to large positive voltages. The data is shown in Fig. \ref{fig:s6} and is well reproduced by our model. The extracted values from this sweep are reported in the main text. The line-by-line fits are shown in Fig. \ref{fig:s7}.

%%%%%%%%%%%%%%%%%%%%%%%%%%%%%%%%%%%%%%%%%%%%%%%%%%%%%%%
%%%%%%%%%%%%%%%%%%%%%%%%%%%%%%%%%%%%%%%%%%%%%%%%%%%%%%

%%%%%%%%%%%%%%%%%%%%%%%%%%%%%%%%%%%%%%%%%%%%%%%%%%%%%%%
%%%%%%%%%%%%%%%%%%%%%%%%%%%%%%%%%%%%%%%%%%%%%%%%%%%%%%

%%%%%%%%%%%%%%%%%%%%%%%%%%%%%%%%%%%%%%%%%%%%%%%%%%%%%%%

%%%%%%%%%%%%%%%%%%%%%%%%%%%%%%%%%%%%%%%%%%%%%%%%

\vspace{-0.4cm}
\section{Electromechanical coupling}
\label{sec:SIelectromech}
\vspace{-0.4cm}

We note that while the piezo-electric effect has a distinct maximum, surprisingly, the electromechanical factor monotonically rises for all observed data. This arises from the relation $k_{33}=d_{33}/\sqrt{s_{33} \epsilon_{r_{33}} \epsilon_{0}}$. Here, because $d\propto \epsilon_r$, the desire for large dielectric constants still hold in general. However, while $d_{33}$ declines with applied bias, the dielectric constant reduces significantly faster, resulting in a higher $k$. With integrated devices with higher fields or strain, higher $k$ should be possible at cryogenic temperatures- unlocking efficient transducers. Even with the limited fields currently, $k^2\approx7\%$ in natural STO, comparable or even exceeding bulk LN.

\vspace{-0.4cm}
\section{Beam steering}
\label{sec:SIsteering}
\vspace{-0.4cm}

%%%%%%%%%%%%%%%%%%%%%%%%%%%%%%%%%%%%%%%%%%%%%%%%%%%%%%%
\begin{figure}[H] 
\begin{center}
\includegraphics[width=1\columnwidth]{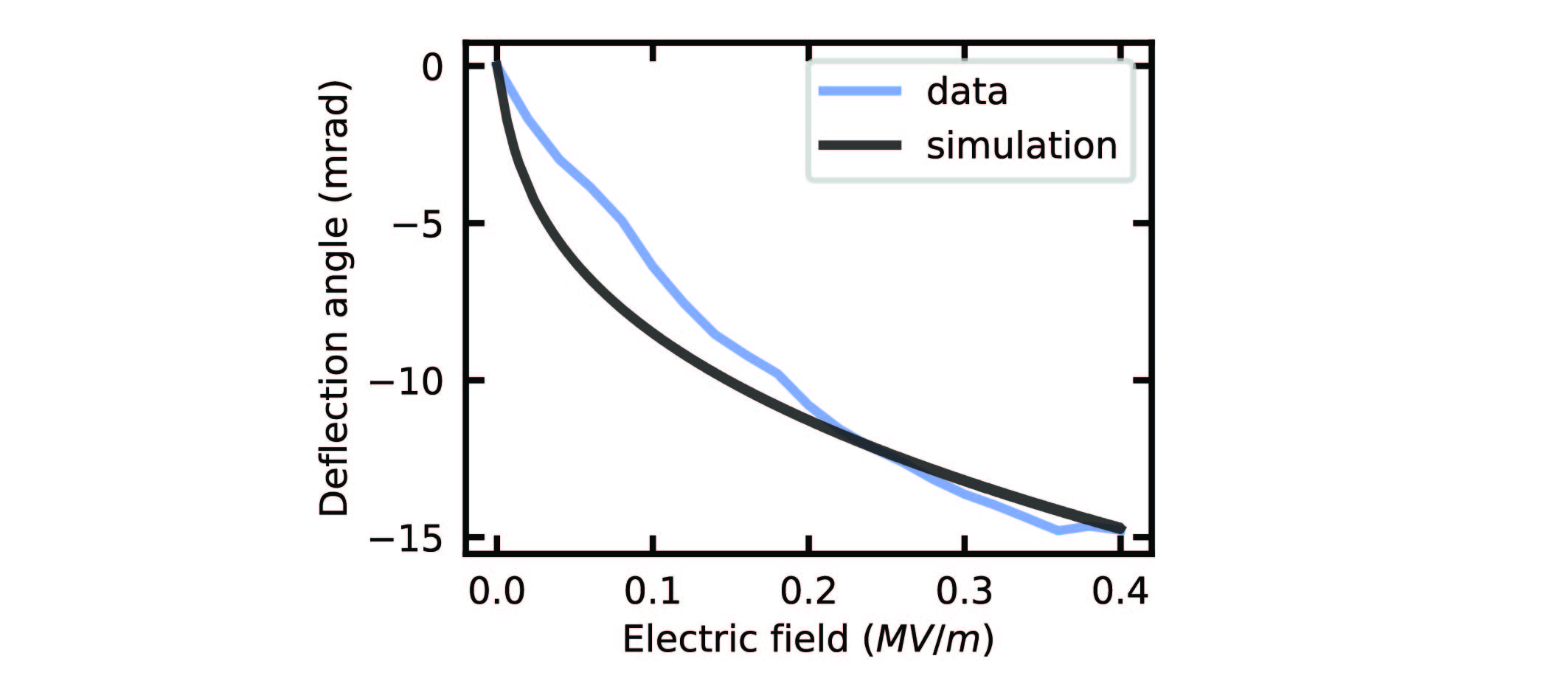}
\caption{
\textbf{Cryogenic beam steering $\vert$} From the quadratic electro-optic effect and trapped charge, we observe large deflection angles as a function of bias. Using the measured field dependent dielectric response from this sample, we can simulate this performance to good agreement assuming a density of trapped charges $\rho=2.7e14$ cm$^{-3}$ and no other free parameters. Data is taken at $T=5$ K.
}
\label{fig:s5}
\end{center}
\end{figure}
%%%%%%%%%%%%%%%%%%%%%%%%%%%%%%%%%%%%%%%%%%%%%%%%%%%%%%

We make another isotopically exchanged STO sample, with an estimated $^{18}$O fraction of 24.5\%. This sample displayed significant beam steering, which we observed with a camera placed after the optical beam passed through the STO chip and the second lens, in Fig. \ref{fig:s4_exp}a. We hypothesize that the origin of the beam steering is oxygen vacancies in the crystal, which trap charges that screen the electric fields applied. While nominally undesirable, trapped charge in KTN EO materials have been shown to make strong field gradients that mediate efficient beam steering. Following the treatment for KTN\cite{Miyazu_2011}, and our measured dielectric response (and therefore electro-optical effect), we can model the observed beam steering with only one free parameter (the charge trap density). Our predicted curve very closely matches the data, shown in Fig. \ref{fig:s5}. Critically, we are able to reproduce the nonlinear nature of the effect due to the bias dependent optical nonlinearity, which is ignored in previous other treatments\cite{Miyazu_2011}.

For comparison, STO is also available in much larger and cheaper crystals compared to KTN for beam steering and other applications. KTN devices would also not work well at cryogenic temperatures due to their near room-temperature phase transition. KTN beam steering has found use in medical devices, Q-switches, ranging systems, and applications that require fast scan speeds using non-mechanical elements  \cite{Chao:17,WANG2023838}. 

\vspace{-0.4cm}
\section{Trends in the $g$ and $q$ tensor elements}
\label{sec:SItrendgq}
\vspace{-0.4cm}

While nominally temperature independent, we show the temperature dependent fit of $g$ and $q$ for all three samples in Fig. \ref{fig:s7}. The dip of the natural STO $q_{11}$ and resulting piezo-electric response at 10-20 K followed by a recovery is potentially due to a temperature-dependent mechanical resonance. We observed in other samples a slight increase in this same temperature range, instead of a dip. The main results of our work are independent of this anomaly. For STO-33, we note distinct increases in $g$ and $q$ from our fit below 15 K (Fig. \ref{fig:s7}). The nature of this increase is potentially related to quantum criticality and requires further investigation.

%%%%%%%%%%%%%%%%%%%%%%%%%%%%%%%%%%%%%%%%%%%%%%%%%%%%%%%
\begin{figure}[htb!] 
\begin{center}
\includegraphics[width=1\columnwidth]{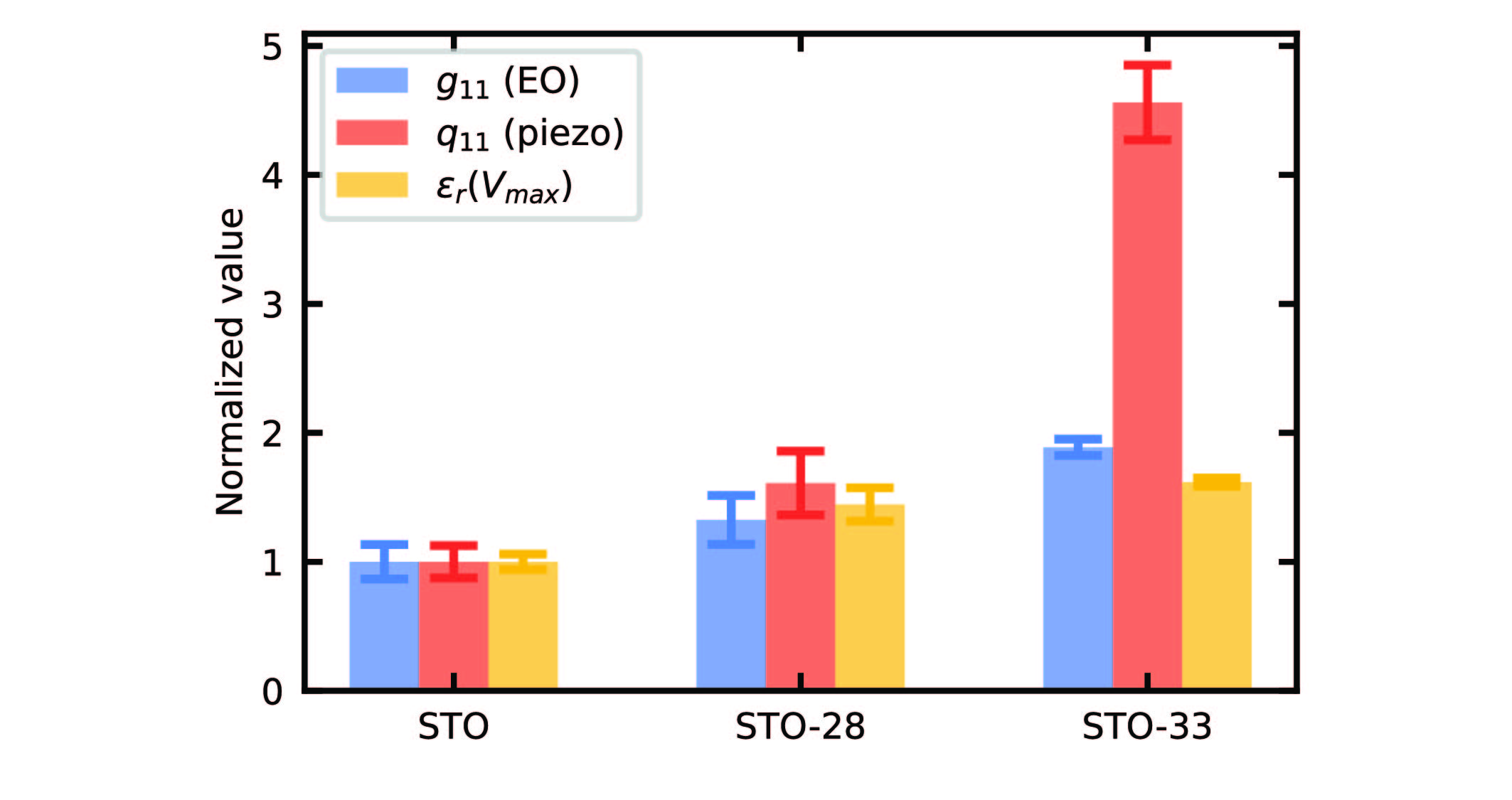}
\caption{ \textbf{Trends in the $g$ and $q$ tensor elements $\vert$} Bar chart of the $T=5$ K $g_{11}$, $q_{11}$ and value of the dielectric constant at the maximum nonlinearity, $\epsilon_{r}(V_{max})$. Values are shown for STO, STO-28, STO-33, normalized to the data for STO.
}
\label{fig:sx}
\end{center}
\end{figure}
%%%%%%%%%%%%%%%%%%%%%%%%%%%%%%%%%%%%%%%%%%%%%%%%%%%%%%

While the observed $g$ tensor has good agreement to previous reports\cite{Fujiicoeff}, the $q$ tensor element is lower by almost exactly a factor of two compared to previous results at room temperature\cite{10.1063/1.4764046}. Changes in the elastic susceptibility $s_{33}$ at low temperatures (not measured here) may drive some of this difference\cite{REHWALD19701483,PhysRevB.61.946}. Despite the strong piezo-electricity, the observed $q_{11}$ is still significantly smaller than predicted with ab-initio theory \cite{10.1063/1.4954886}

The trends of the $g$ and $q$ tensor values between samples are shown in Fig. \ref{fig:sx}. The relative increase in STO-28 is the same for both electro-optic and piezo-electric nonlinearity, but near quantum criticality, $q$ changes much more dramatically. The observed $g$, $q$ also both increase with increases in the dielectric constant at the maximum value of $P\epsilon_r$. We note that this is drastically different from traditional ferroelectrics, where $g$ remains nearly constant across all materials, and $q$ varies inversely with dielectric constant \cite{uchino}. In fact, near quantum criticality, we observe an electro-optic $g$ nearly three times larger than any other perovskite.

\vspace{-0.4cm}
\section{Comparison with other nonlinear materials}
\label{sec:SIcomparison}
\vspace{-0.4cm}

In Fig. \ref{fig:fig4} of the main text, a comparison between nonlinear materials is displayed. There are variances in experimental reports and it is difficult overall to match dielectric and nonlinear measurements to the same source. In general, we use: `Oxides’ of Volume 36 `Ferroelectrics and Related Substances’ of Landolt-Börnstein - Group III Condensed Matter."\cite{landolt}, otherwise more references follow:

\begin{center}
   LiNbO$_3$\cite{LN, Boydtextbook,Hamze2020}\\
    LiTaO$_3$\cite{LT,Hamze2020}\\
    PbTiO3$_3$\cite{PTO1,PTO2}\\
    BaTiO$_3$\cite{BTO,Hamze2020}\\
    KNbO$_3$\cite{KNO,Hamze2020}\\
    PZT (various compositions)\cite{LIU2023135,PZT}\\
    PMN-PT (various compositions, including PIN-PMN-PT)\cite{LIU2023135,doi:10.1126/science.abn7711}\\
    KTN (various compositions)\cite{KTN1,KTN2}\\ 
\end{center}
%KTN EO reference has nice enahncement near Tc, kTN1

\vspace{-0.4cm}
\section{Trends and comparisons in nonlinear materials}
\vspace{-0.4cm}

\vspace{0 cm}
\subsection*{Nonlinearities as a function of dielectric constant}
\vspace{-0.4cm}

%%%%%%%%%%%%%%%%%%%%%%%%%%%%%%%%%%%%%%%%%%%%%%%%%%%%%%%
\begin{figure}[htb!] 
\begin{center}
\includegraphics[width=1\columnwidth]{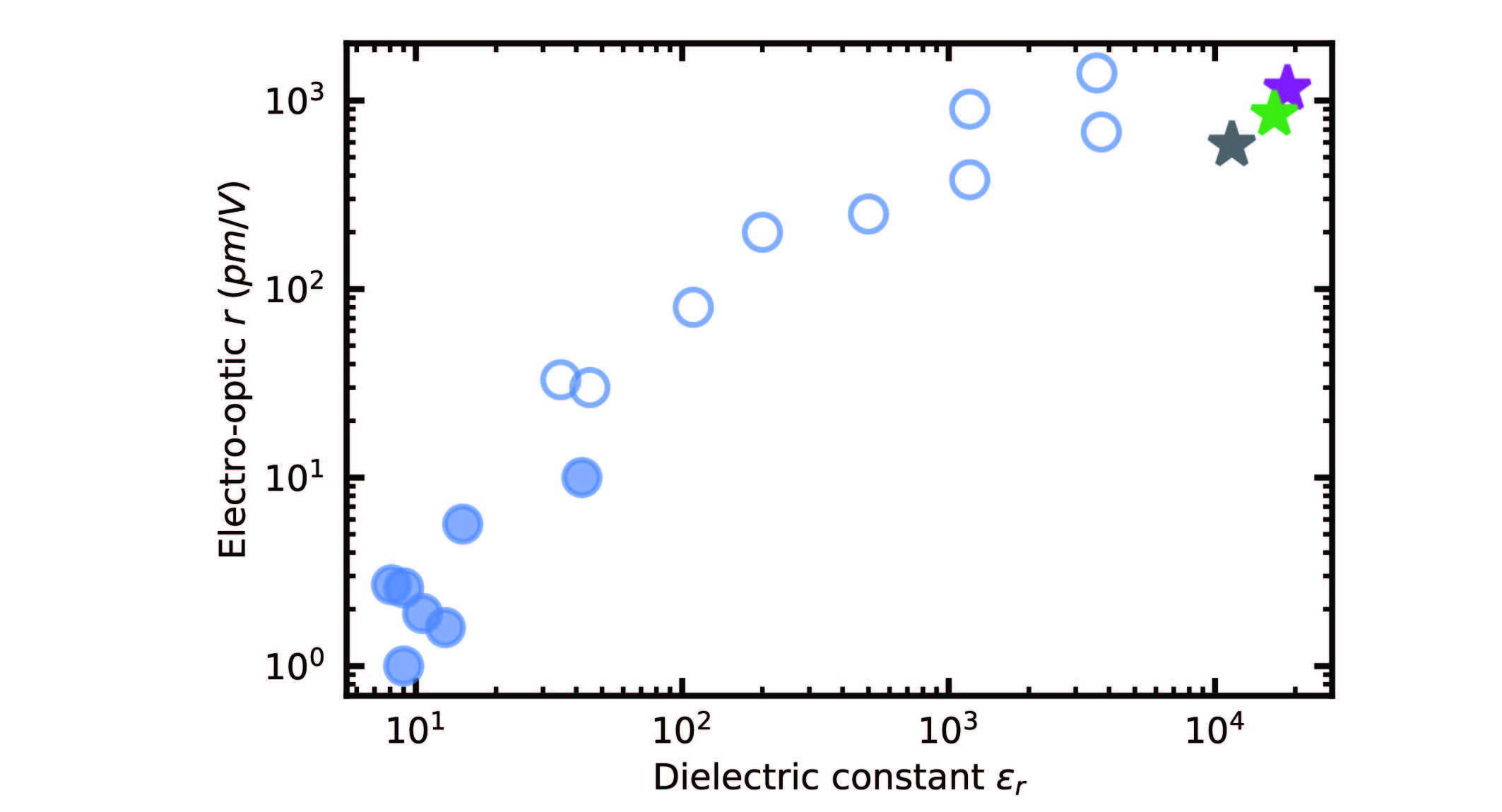}
\caption{
\textbf{Scaling of the electro-optic effect with dielectric constant $\vert$} \textbf{a}, Log-log plot of various electro-optic materials, including non perovskites (filled circles). The starred data points are STO at $T=5$ K, following the labeling in the main text.
}
\label{fig:s11}
\end{center}
\end{figure}
%%%%%%%%%%%%%%%%%%%%%%%%%%%%%%%%%%%%%%%%%%%%%%%%%%%%%%

\begin{figure}[htb!] 
\begin{center}
\includegraphics[width=1\columnwidth]{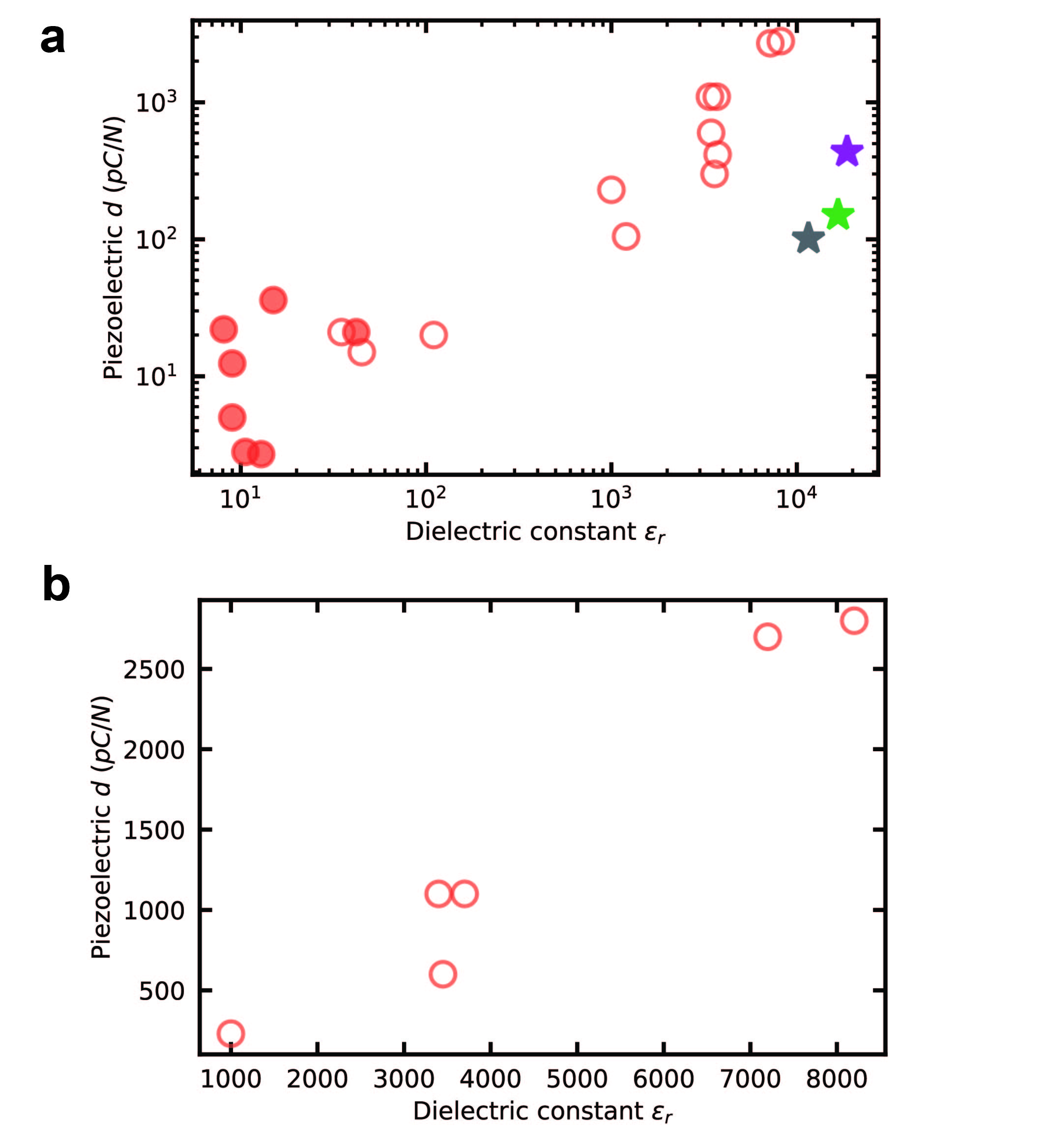}
\caption{
\textbf{Scaling of the piezo-electric effect with dielectric constant $\vert$} \textbf{a}, Log-log plot of various piezo-electric materials, including non perovskites (filled circles). The starred data points are STO at $T=5~K$, following the labeling in the main text. \textbf{b}, Piezo-electric coefficients of a series of lead-zirconate-titanate materials, including PIN-PMN-PT, PZT, PMN-PT, as a function of dielectric constant.
}
\label{fig:s10}
\end{center}
\end{figure}

We also evaluate the scaling of piezo-electric and electro-optic nonlinearities with \textit{just} the dielectric constant, shown in Figures \ref{fig:s11}, \ref{fig:s10}. This includes non ferroelectric materials like AlN, GaN, GaAs, KDP, KTP, BBO, ZnO, etc (where there is no polarization in the same sense as in ferroelectrics). We see that the dielectric constant is still a good metric for materials discovery and guidance. However, the trend is not purely linear in the electro-optic case, Fig. \ref{fig:s11}. For the piezo-electric case, the trend is more complicated (Fig. \ref{fig:s10}a), most likely due to non ``ionic" contributions to the piezo-electric tensor. On the other hand Fig. \ref{fig:s10}b shows the trend for a class of similar lead-zirconate-titanate compounds, showing good linearity. Therefore, dielectric constant is still a fair metric to identify nonlinear materials.

%%%%%%%%%%%%%%%%%%%%%%%%%%%%%%%%%%%%%%%%%%%%%%%%%%%%%%%
%%%%%%%%%%%%%%%%%%%%%%%%%%%%%%%%%%%%%%%%%%%%%%%%%%%%%%

\vspace{-0.4cm}
\subsection*{Comparison at cryogenic temperatures of $n^3r$}
\vspace{-0.4cm}

\begin{figure}[htb] 
\begin{center}
\includegraphics[width=1\columnwidth]{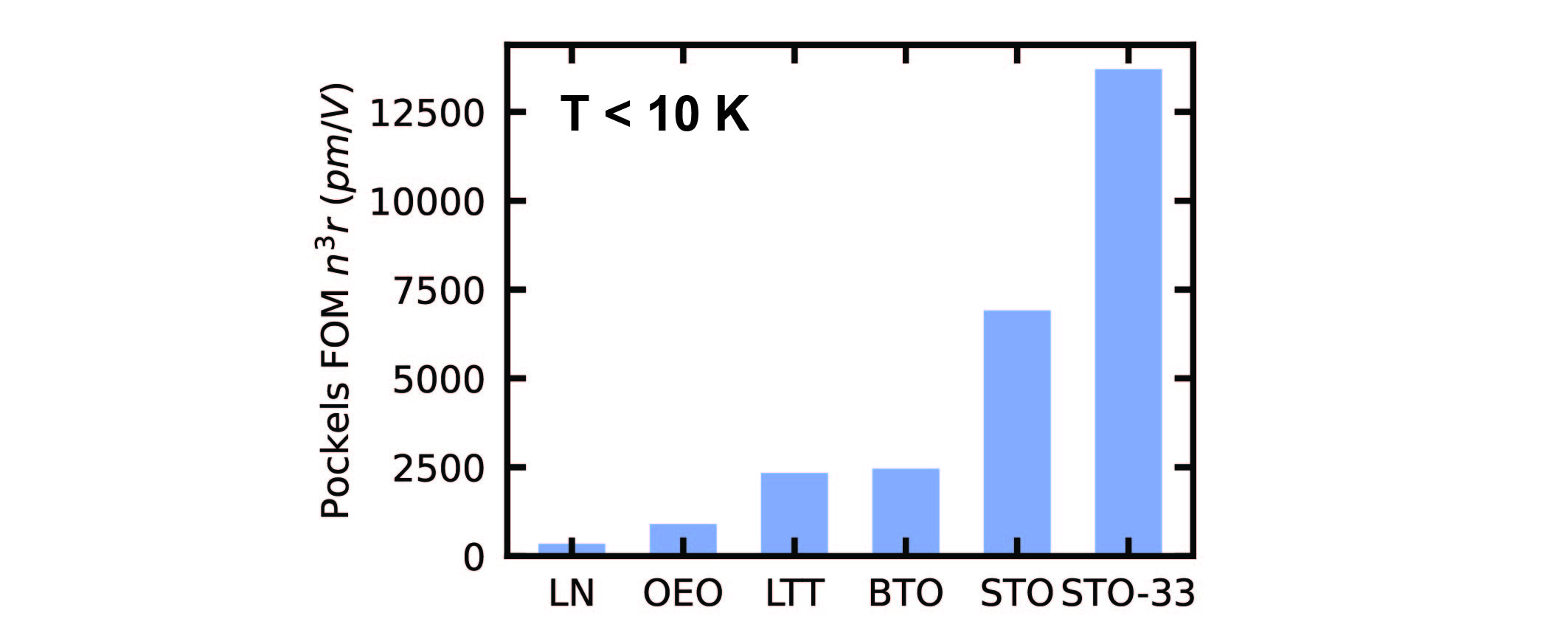}
\caption{
\textbf{Comparison of cryogenic electro-optic materials $\vert$} Pockels FOM $n^3r$ for various materials below 10 K.
}
\label{fig:scryo}
\end{center}
\end{figure}

In Fig. \ref{fig:scryo} we plot the reported Pockels figure of merit, $n^3r$, for materials with reported values at 10 K or lower. LTT is lithium thallium tartrate monohydrate, a known low $T_c$ ferroelectric \cite{LTTEO}. OEO stands for organic-electro-optic polymer.

Fig. \ref{fig:fig4}b in the main text similarly shows a comparison of the piezo-electric coefficients at 10 K and lower. Values of various lead-zirconate-titanate compositions are similar at these low temperatures \cite{Cryopiezo}. As LN is far away from $T_c$ at both 297 K and below 10 K, its nonlinearities do not change significantly when cooled.

%%%%%%%%%%%%%%%%%%%%%%%%%%%%%%%%%%%%%%%%%%%%%%%%%%%%%%%
%%%%%%%%%%%%%%%%%%%%%%%%%%%%%%%%%%%%%%%%%%%%%%%%%%%%%%

\section{Outlook for quantum and classical systems}
\label{sec:SIoutlook}
\vspace{-0.4cm}

For quantum systems utilizing electro-optics or piezo-electricity, the relevant parameter is the coupling coefficient $g$ (not the same as $g_{11}$). $g\propto r,d$ but also the impedance (or equivalently, the size/capacitance) of the device $\propto 1/\sqrt{\epsilon}$. Because $r,d$ are proportional to $\epsilon$, then we can estimate $g\propto\sqrt{\epsilon}$. Another metric is the cooperativity ($C$) between optical(mechanical) and electrical subsystems, $C=g^2/(\kappa \gamma)$. Unfortunately, as the dielectric constant grows, so does the electrical loss. If we assume the imaginary dielectric response (loss) scales similarly to the dielectric constant, we can estimate that $C$ does not scale with dielectric constant of the material. Therefore, we propose that materials engineering primarily improves $g$, but not $C$. However, for certain applications, large $g$ is desirable specifically. For electro-optic quantum transduction, the important consideration is to reduce pump induced heating, where the relevant figure-of-merit scales with $g^2$, and does not depend on the dielectric loss \cite{Jiang2020}. With STO, similar transduction efficiencies can be achieved with at least 400 times less power compared to LN.

Cryogenic electro-optic devices also offers an attractive alternative solution to the “wiring problem” in superconducting systems, as optical fiber would allow for millions of signals to reach millikelvin environments with minimal heating\cite{Pintus2022,lecocq2021control}. Finally, For photonic switches, STO would allow at least a 5x reduction in modulator size at 5 K due to increased $r$. This is critical as loss is nonlinear with device length.

On the other hand, it is desirable to also minimize the electrical power consumption per bit $E_{bit}=\frac{1}{4}CV^2$ of a potential cryogenic optical modulator. The required voltage scales with 1/$(L r)$, where L is the length of the modulator. On the other hand, the capacitance $C$ scales with L and $\epsilon$, so $E_{bit}\propto \epsilon/(L r^2)$. Therefore, one can estimate that materials have $E_{bit}\propto 1/\epsilon$, if $r\propto \epsilon$ for a given device size. Higher dielectric constant materials closer to phase transitions therefore show improvement in classical modulator application. Other example applications of cryogenic systems can be found in the supplemental information of related work on BTO modulators\cite{Eltes2020}.

\vspace{-0.4cm}
\section{Summary of $^{18}$O isotope exchange and annealing}
\label{sec:SI18O}
\vspace{-0.4cm}

As explained in the Methods section, we perform nanoscale-secondary-ion-mass-spectroscopy (nanoSIMS) to determine the relative ratio of $^{16}$O to $^{18}$O. We also compare this with the expected $^{18}$O ratio calculated by measuring the mass of the STO before the $^{18}$O rich annealing steps and after. Below, we summarize these results:

\begin{table}[H]
\vspace{-0.25cm}
\caption{\label{tab:ISO} \textbf{$^{18}$O concentrations and total duration in days of annealing for the SrTiO$_3$ samples}}
\begin{ruledtabular}
\begin{tabular}{@{\extracolsep{\fill}}lp{1.5cm}p{1.5cm}p{1.5cm}p{2.5cm}}
\textbf{} & \textbf{STO \newline (Natural)} & \textbf{STO-28} & \textbf{STO-33} & \textbf{Beam \newline Steering} \\ 
\hline 
nanoSIMS & $0.22(4)\% $ \newline & $28.3(3)\%$ & $33.0(5)\% $ & - \\
Weight-based &  N/A & $35.92\%$ \newline & $36.4\%$ & $24.48\%$ \\
$^{18}$O anneal & 0 & 27.8\newline & 48.8 & 27.8 \\
Vacuum anneal  & 0 & 27.8 & 41.8 & 27.8 \\

\end{tabular}
\end{ruledtabular}
\end{table}

We report on the nanoSIMS values in the main text, as they do not depend on other sources of mass change which confound a purely mass based measurement.

\clearpage

% \bibliography{main}
\putbib[main]  
\end{bibunit}

\end{document}